\documentclass[10pt]{article}
\usepackage{amsmath}
\usepackage{graphicx}
\usepackage{amsfonts}
\usepackage{amssymb}
\usepackage{epsf}
\usepackage{latexsym}

\def\80{\hspace{0.8in}}

\def\nm{\mbox{m}}

\newcommand{\be}{\begin{enumerate}}
\newcommand{\ee}{\end{enumerate}}
\newcommand{\bi}{\begin{itemize}}
\newcommand{\ei}{\end{itemize}}
\newcommand{\bd}{\begin{description}}
\newcommand{\ed}{\end{description}}
\def\beq{\begin{equation}}
\def\eeq{\end{equation}}
\def\bea{\begin{eqnarray}}
\def\eea{\end{eqnarray}}
\def\hat{\widehat}

\def\half{\frac{1}{2}}
\def\pa{\partial}
\def\d{\textrm{d}}
\def\Rec{\mbox{\textcircled{$\star$}}}        

\def\Spade{\mbox{$\spadesuit$}}           
\def\Club{\mbox{$\clubsuit$}}             

\def\bupSigma{\mbox{\boldmath$\Sigma$}}      

\def\lt{\mbox{\Large $t$}}

\def\NSI{Na\"{\i}ve Schr\"{o}dinger Interpretation }

\def\Star{\mbox{\Large$\ast$}}                

\def\Bigalpha{\mbox{\Large $\alpha$}}         
\def\Bigbeta{\mbox{\Large $\beta$}} 
\def\bigalpha{\mbox{\normalsize $\alpha$}}        
\def\bigbeta{\mbox{\normalsize $\beta$}} 

\def\mc{\mbox{c}}
\def\md{\mbox{d}} 
\def\me{\mbox{e}}

\def\mh{\mbox{h}}
\def\mi{\mbox{i}}
\def\mj{\mbox{j}} 

\def\ml{\mbox{l}}

\def\nk{\mbox{k}}
\def\mm{\mbox{m}}
\def\mn{\mbox{n}}

\def\mp{\mbox{p}} 

\def\nm{\mbox{m}}

\def\nr{\mbox{r}}
\def\ms{\mbox{s}}

\def\mC{\mbox{C}}
\def\mD{\mbox{D}}

\def\mF{\mbox{F}}

\def\mH{\mbox{H}} 
\def\mI{\mbox{I}}
\def\mJ{\mbox{J}}

\def\mL{\mbox{L}}

\def\mN{\mbox{N}} 

\def\mP{\mbox{P}}

\def\mR{\mbox{R}}
\def\mS{\mbox{S}}
\def\mT{\mbox{T}}

\def\mY{\mbox{Y}}
\def\mZ{\mbox{Z}}

\def\sa{\mbox{\scriptsize a}}

\def\sc{\mbox{\scriptsize c}}
\def\sd{\mbox{\scriptsize d}}
\def\se{\mbox{\scriptsize e}}

\def\sh{\mbox{\scriptsize h}} 
\def\si{\mbox{\scriptsize i}}
\def\sj{\mbox{\scriptsize j}} 
\def\sk{\mbox{\scriptsize k}}
\def\sll{\mbox{\scriptsize l}}  
\def\sm{\mbox{\scriptsize m}}
\def\sn{\mbox{\scriptsize n}} 
\def\so{\mbox{\scriptsize o}} 
\def\sp{\mbox{\scriptsize p}}

\def\sr{\mbox{\scriptsize r}}
\def\sss{\mbox{\scriptsize s}}
\def\st{\mbox{\scriptsize t}}

\def\sB{\mbox{\scriptsize B}}
\def\sC{\mbox{\scriptsize C}}
\def\sD{\mbox{\scriptsize D}}

\def\sG{\mbox{\scriptsize G}}

\def\sI{\mbox{\scriptsize I}}
\def\sJ{\mbox{\scriptsize J}}
\def\sK{\mbox{\scriptsize K}}

\def\sN{\mbox{\scriptsize N}}

\def\sR{\mbox{\scriptsize R}}
\def\sS{\mbox{\scriptsize S}}
\def\sT{\mbox{\scriptsize T}}
\def\sU{\mbox{\scriptsize U}}

\def\sW{\mbox{\scriptsize W}}
 
\def\sY{\mbox{\scriptsize Y}}

\def\Rec{\mbox{\textcircled{$\star$}}}        
\def\Spade{\mbox{$\spadesuit$}}           
\def\Club{\mbox{$\clubsuit$}}             

\def\eph(B){\mbox{\scriptsize emergent(LMB)}}

\def\fs{\mbox{\sffamily s}}

\def\fE{\mbox{\sffamily E}}

\def\fG{\mbox{\sffamily G}}

\def\fM{\mbox{\sffamily M}}
\def\fN{\mbox{\sffamily N}}

\def\fQ{\mbox{\sffamily Q}}

\def\fS{\mbox{\sffamily S}}

\def\fV{\mbox{\sffamily V}}
\def\fW{\mbox{\sffamily W}}

\def\sfa{\mbox{\sffamily{\scriptsize a}}}

\def\sfA{\mbox{\sffamily{\scriptsize A}}}
\def\sfB{\mbox{\sffamily{\scriptsize B}}}
\def\sfC{\mbox{\sffamily{\scriptsize C}}}

\def\sfM{\mbox{\sffamily{\scriptsize M}}}

\def\sfZ{\mbox{\sffamily{\scriptsize Z}}}
\def\K{Kucha\v{r} }

\def\FA{\mbox{\Large $\mathfrak{a}$}}

\def\FC{\mbox{\Large $\mathfrak{c}$}}

\begin{document}

\begin{titlepage}

\begin{center}

{\LARGE\bf RELATIONAL QUADRILATERALLAND. II}

\vspace{.1in}

{\LARGE\bf THE QUANTUM THEORY}

\vspace{.2in}

{\bf Edward Anderson}$^1$ and {\bf Sophie Kneller}$^2$

\vspace{.2in}

\large {\em $^1$ DAMTP Cambridge} \normalsize

\large {\em $^2$ Murray Edwards College, Cambridge}

\vspace{.2in}

\end{center}

\begin{abstract}

This paper provides the quantum treatment of the relational quadrilateral. 
The underlying reduced configuration spaces are $\mathbb{CP}^2$ and the cone over this, $\mC(\mathbb{CP}^2)$. 
We consider exact free and isotropic HO potential cases and perturbations about these.
Moreover, our purely relational kinematical quantization is distinct from the usual one for $\mathbb{CP}^2$, which turns out to carry absolutist connotations instead.  
Thus this paper is the first to note absolute-versus-relational motion distinctions at the kinematical rather than dynamical level. 
It is also an example of value to the discussion of kinematical quantization along the lines of Isham 1984.  
This treatment of the relational quadrilateral is the first relational QM with very new mathematics for a finite QM model. 
It is far more typical of the general quantum relational $N$-a-gon than the previously-studied case of the relational triangle.  
We consider useful integrals as regards perturbation theory and the peaking interpretation of quantum cosmology.  
We subsequently consider problem of time applications of this: quantum \K beables, the Machian version of the semiclassical approach 
and the timeless na\"{\i}ve Schr\"{o}dinger interpretation.
These go toward extending the combined Machian semiclassical-Histories-Timeless Approach of [1] to the case of the quadrilateral, which will be treated in subsequent papers.  

\end{abstract}

\vspace{0.2in} 


\vspace{0.2in}  

\mbox{ }

\mbox{ }   

\vspace{3in}

\mbox{ } 

\noindent $^1$ Corresponding Author:  ea212@cam.ac.uk 

\end{titlepage}

\section{Introduction}\label{Intro2}

The present paper considers the quantum counterpart of Paper I's \cite{QuadI} classical work on the quadrilateralland relational particle model (RPM) \cite{BB82, B03}.\footnote{Refer 
to Paper I for details of motivation for RPM's in general and quadrilateralland in particular, for notation and for equation/section/figure references that begin with `I.'}
%
Quantum RPM's were first considered by Julian Barbour, Lee Smolin and Carlo Rovelli \cite{BSSmolin, Rovelli}, though practical progress with solving concrete examples of these was hindered until 
\cite{Paris, 06I, 06II}. 
See \cite{SemiclI, etc, Records, 08I, 08II, AF, +tri, MGM, ScaleQM, 08III,  APoT, SemiclIII, QuadI, AHall, ARel, APoT2, ML12, FileR, BKM13, ACos2} for subsequent development.
This sudden progress was triggered by 1) Barbour formulating the pure-shape RPM \cite{B03, Piombino}.  
This proved to be easier to solve. 2) 
By the sequence of Keys and profitable interdisciplinary observations given in Paper I.  
(\cite{Smith60, Dragt, Kuiper, GiPo, Kendall8489, LR97, Clemence, Kendall, MacFarlane} are particular antecedents for this  from Molecular Physics, Celestial Mechanics, Geometrical Methods 
and Shape Statistics).

RPM's have been valued as model arenas \cite{Kuchar92, EOT, Kieferbook, APoT, B11GrybTh, Gryb1, ARel, APoT2, FileR, ML12, BKM13} for the Problem of Time (PoT) 
\cite{Kuchar92, I93, APoT2} in Quantum Gravity via the analogies exposited in \cite{RWR, FileR}.  
They have been used e.g. for study of i)Timeless Approaches \cite{B94III, EOT, Records, AF, +tri, ScaleQM, 08III, ARel2, FileR}. 
ii) The Semiclassical Approach \cite{SemiclI, ScaleQM, 08III, SemiclIII, FileR, ACos2} in close parallel to Halliwell--Hawking's approach \cite{HallHaw} to Quantum Cosmology.
iii) Histories Theory and its combination with the previous two \cite{AHall, FileR}.

In more detail, as argued in \cite{APoT2}, the Temporal Relationalism and Configurational Relationalism which RPM's are constructed to embody are 2 of the 8 PoT facets.
Which classical PoT aspects are present for RPM's and how to resolve them was covered in Paper I.  
At the quantum level, if Configurational Relationalism was resolved at the classical level, it stays resolved. 
Temporal Relationalism now resurfaces as the Frozen Formalism Problem; various strategies for this were laid out in Paper I.  
The main one followed in the present paper is the Semiclassical Approach \cite{FileR, ACos2}.  
The Problem of Beables was resolved at the level of classical \K beables in Paper I. 
The present paper considers the quantum counterpart of this.  
As regards other PoT facets, the following statement holds at both the classical and quantum levels. 
RPM's are conceptually free of the Foliation Dependence and Spacetime Reconstruction Problems, and can be cast as free from the Constraint Closure Problem too.  
That covers the 6/8ths of the PoT required to have a local resolution, with two caveats that require further papers (III and IV). 

\noindent 1) The Semiclassical Approach requires support from other approaches, in particular Histories Theory and Timeless Records \cite{QuadIV, QuadV}.  

\noindent 2) Quantum Dirac beables have not yet been considered for quadrilateralland; the current program's \cite{FileR} way of addressing these 
uses the machinery of 1) and thus must await treatment of that.  

\noindent RPM's and similar are also often used to motivate the Linking Theory Approach to Shape Dynamics \cite{B11GrybTh, GGKM11, BKM13}. 
Other papers found uses in investigating such as whole-universe path integral approaches \cite{Gryb1}, geometrical quantization \cite{Rovelli}, 
operator-ordering in Quantum Cosmology \cite{Banal, FileR}, and an investigation of an alternative anomaly-based emergent-time mechanism \cite{ML12}.
%

Quantum RPM's hitherto studied are scaled 3-stop metroland \cite{06I, SemiclI, MGM, SemiclIII}, pure-shape 4-stop metroland \cite{AF} pure-shape triangleland QM \cite{08II, +tri}, 
scaled 4- and $N$-stop metroland \cite{ScaleQM}, and scaled triangleland \cite{08III, SemiclIII, AHall}. 
See Part III of \cite{FileR} for a review of these.
Quadrilateralland QM (pure-shape and scaled) is then logically the next step for this program and that taken here in the present article. 
Compared to the preceding list, there is now nontrivial $\mathbb{CP}^k \leftrightarrow SU(k + 1)$ mathematics to contend with.  
The $N$-a-gon is unlikely to prove much harder than the quadrilateral. 
On the other hand, the triangle is exceptionally simpler by benefitting in non-generalizable ways because its $\mathbb{CP}^1 = \mathbb{S}^2$ allows for extra techniques.

\mbox{ }

\noindent Some interdisciplinary comments on the present Paper are as follows.  
\cite{NdH, Norway} consider the atom in $N$-$d$ (in the sense of a 1/$r$ potential in dimension $N$). 
\cite{Norway} considered the Stark effect not only for $N$-$d$ atomic models but for $N$-$d$ rotors as well. 
Both of these are maximally symmetric problems (on $\mathbb{R}^p$ and $\mathbb{S}^p$, each of which possess p\{p + 1\}/2 Killing vectors).
MacFarlane's work \cite{MacFarlane} and the current paper can then be viewed as an extension of this work for the next most symmetric case of $\mathbb{CP}^2$ that exists 
for shape space dimension $q = 4$. 
(This has 8 Killing vectors rather than the maximal 10.)  
Our paper's useful integrals for QM on $\mathbb{CP}^2$ further extend MacFarlane's work to perturbations about the free case.
See \cite{CPNHO} for other literature concerning HO's on $\mathbb{CP}^N$, though we do not know of any previous literature that covers the $\mathbb{CP}^2$ counterpart of the Stark effect.
It is nontrivial as a robustness test of the atom, in that it unveils a number of fortunate occurrences for the standard orbitals and Stark effect 
that end upon passing from maximal to the next most maximal symmetry.  
See the Conclusion for examples of this.  
One part of the interpretation of QM of a quadrilateralland involves an application of Paper I's complex-projective chopping board counterpart of Kendall's spherical blackboard from 
Shape Statistics \cite{Kendall8489}.  
We shall also see that this QM is a cross between the Periodic Table and Gell-Mann's eightfold way from Particle Physics, in a sense made precise in Secs \ref{free-solve} and \ref{vis}.  
We shed light on how quadrilateralland's HO-type systems are far more like triangleland's than 4-stop metroland's at the quantum level.  
This is despite their greater classical similarity with 4-stop metroland indicated in Sec I.25.  
This has further relevance as regards `triangleland within quadrilateralland' robustness tests paralleling \K and Ryan's work \cite{KR89} in minisuperspace Quantum Cosmology. 
%

\mbox{ } 

\noindent An outline of this Paper is as follows. 
In Sec \ref{KinQuant}, we consider kinematical quantization for pure-shape quadrilateralland.
It is an interesting example as regards Isham 1984 kinematical quantization \cite{I84} and as regards how the absolute versus relational motion debate already shows up at the level of 
kinematical quantization. 
(One of us previously pointed out distinctions of this type at the subsequent level of the wave equations themselves \cite{FileR}.)  
In Sec \ref{TISE}, we construct the conformal-ordered TISE for pure-shape quadrilateralland, which we separate for the free case in Sec \ref{sep} in Gibbons--Pope type coordinates 
\cite{GiPo, MacFarlane}. 
Sec \ref{free-solve} covers the energies, quantum numbers and wavefunctions for this problem and Sec \ref{vis} describes the ground state and first few excited states.
Sec \ref{scale} considers scaled quadrilateralland, in particular for isotropic HO's.  
Sec \ref{Use-Int} constructs useful integrals out of the scaled isotropic HO and pure-shape free wavefunctions (these generalize the integrals used in e.g. the study of the Stark effect).
These are then used in Sec \ref{Exp} for peak and spread analysis (`Peaking Interpretation of Quantum Cosmology', though we show this has a well-known counterpart in Atomic Physics). 
Thay are also used in Sec \ref{Semicl} for time-dependent perturbation theory in the Semiclassical Quantum Cosmology analogue model context.  
Sec \ref{NSI} finishes Paper I's consideration of \NSI questions for quadrilateralland; this approach is a prequel to the Machian version of Halliwell's 
combined approach's use of regions in Paper IV. 
We conclude in Sec \ref{Concl2}, including a sketch of extensions to the general $N$-a-gonland.

\section{Kinematical quantization of quadrilateralland}\label{KinQuant} 

In general, one has to make a {\it choice} \cite{I84} of a preferred subalgebra of functions of one's configurations $Q^{\sfC}$ and momenta $P_{\sfC}$ 
that are the ones to be promoted to QM operators.
Some context for this is that the {\sl Groenewold--van Hove phenomenon} \cite{Gotay} precludes simultaneous promotion of all classical quantities to quantum operators.  
There are also global considerations \cite{I84} by which a model's quantum commutator algebra is not in general isomorphic to that problem's classical Poisson bracket algebra.

\noindent\underline{Key 18} The RPM program lies within Isham's \cite{I84} $\fQ/\fG$ example for $\fG$ a subgroup of $\fQ$. 
Then the relevant spaces involved in kinematical quantization can be decomposed as semisimple products $\fV^*(\fQ) \, \mbox{\textcircled{S}} \, \fG_{\sc\sa\sn}(\fQ)$. 
Here, $\fG_{\sc\sa\sn}(\fQ)$ is the canonical group and $\fV^*$ is the dual of a linear space $\fV$ that is natural due to its carrying a linear representation of 
$\fQ$ that realizes the $\fQ$ orbits.
Mackey Theory \cite{I84} is then a powerful tool for finding the representations of such semidirect product algebras.  
Furthermore, $\fV^* = \fV$ for finite examples and $\fG_{\sc\sa\sn}(\fQ)$ = Isom($\fQ$) for all 1- and 2-$d$ RPM's.  

\mbox{ }  

\noindent Example 0) For absolute $\mathbb{R}^p$, the canonical group is Isom($\mathbb{R}^p$) = $\mbox{Eucl}(p) = \mbox{Tr}(p) \, \mbox{\textcircled{S}} \, \mbox{Rot}(p) 
                                                                                                       = \mathbb{R}^{p} \, \mbox{\textcircled{S}} \, SO(p)$.

Then an appropriate linear space is $\mathbb{R}^{p}$, so, overall, one has  $\mathbb{R}^n \, \mbox{\textcircled{S}} \, \mathbb{R}^{n} \, \mbox{\textcircled{S}} \, SO(n)$.
These are the $x^i$, their conjugates the $p_i$ and the corresponding angular momenta $L_i = \epsilon_{ij}\mbox{}^kx^jp_k$, 
so this case is both physically and mathematically very familiar.   

\noindent Example 1) For scaled $N$-stop metroland,  ${\cal R}(N, 1) = \mathbb{R}^n$, so the outcome is mathematically the same as above. 
However, physically the roles of the objects involved are relative Jacobi separations $\rho^i$, their conjugates $\pi_i$ and relative dilational momenta 
${\cal D}\mi\ml_{\Gamma}$ (Sec I.18) for $\Gamma$ running over $SO(n)$'s 1 to $n\{n - 1\}/2$ indices.

\noindent Example 2) For scalefree $N$-stop metroland, $\fS(N, 1) = \mathbb{S}^{n - 1}$, for which the canonical group is Isom($\mathbb{S}^{n - 1}) =\mbox{Rot}(n) = SO(n)$.  
Then an appropriate linear space is $\mathbb{R}^n$. 
Now the objects in question are the ${\cal D}\mi\ml_{\Gamma}$  again, alongside the $n^i$ that square to 1 so as to provide the on-$\mathbb{S}^{n - 1}$ condition.  
These unit Cartesian vectors in configuration space are most conveniently expressed in ultraspherical coordinates (since the ${\cal D}\mi\ml_{\Gamma}$ are).  

\noindent Example 3) For pure-shape $N$-a-gonland's $\fS(N, 2) = \mathbb{CP}^{n - 1}$ shape space, the canonical group is $\fG_{\sc\sa\sn}(\fS(N, 2)) =$

\noindent $\mbox{Isom}(\mathbb{CP}^{n - 1}) = SU(n)/\mathbb{Z}_n$ 
Moreover, this shape space can also be written as $SU$($n$)/$U$($n$ -- 1); thus it is also a subcase of the general form in Isham's example above. 

\noindent Then one possible kinematical quantization involves $\fV \, \mbox{\textcircled{S}} \, \fG_{\sc\sa\sn} = SU(n)/\mathbb{Z}_n \, \mbox{\textcircled{S}} \, \mathbb{R}^{2n}$,  
for $\mathbb{R}^{2n}$ better thought of as  $\mathbb{C}^{n}$ \cite{I84}.  

\mbox{ } 

\noindent Note however that triangleland admits a distinct kinematical quantization. 
I.e. $\fV \, \mbox{\textcircled{S}} \, \fG_{\sc\sa\sn} = SO(3) \, \mbox{\textcircled{S}} \, \mathbb{R}^3$ with the $\mathbb{R}^3$ made up of the Dragt coordinates \cite{Dragt}, (I.31) 
$\mbox{Dra}^{\Gamma}$ and the SO(3) of mixed relative angular momentum and relative dilational momentum quantities as per Sec I.18.    
Moreover, this alternative i) does not involve postulating objective existence to absolute entities (present among the $\mathbb{C}^2$ of relative Jacobi vectors) and ii) 
is a more minimal realization (3-$d$ to 4-$d$).

Moreover, generalizing the latter relational kinematical quantization of the triangle to the quadrilateral is not particularly obvious.   
Progress can be made via noting that $\mathbb{R}^3$ is also IHP$(\mathbb{C}^2$, 2) -- i.e. the space of irreducible homogeneous polynomials of degree 2 (Sec I.16) -- 
via the 3-vector to Pauli matrix map that rests on the well-known accidental relation between $SU(2)$ and $SO(3)$.    
IHP$(\mathbb{C}^{n}, 2)$ then continues to be available for general-$n$ $\mathbb{CP}^{n - 1}$ kinematical quantization. 
In the quadrilateralland case, this space is composed of the 8 independent shape quantities of Sec I.13.
All in all, we have the kinematical quantization $\fV \, \, \mbox{\textcircled{S}} \,  \, \fG_{\sc\sa\sn} = SU(3) \, \, \mbox{\textcircled{S}} \, \, \mI\mH\mP(\mathbb{C}^n, 2)$
This is {\sl not} the minimal-sized space for any $N > 3$ since $2n < n^2 - 1$ for all integer $n > 2$.  
Nevertheless, minimality is {\sl a guideline} and not an obligation, and argument i) continues to stand.  

\mbox{ } 

\noindent Next, the quadrilateralland isometry generators are $T_{\Gamma} = 
\{\widehat{\cal Y}, \widehat{\cal I}_3, \widehat{\cal I}_{\pm}, \widehat{\cal U}_{\pm}, \widehat{\cal V}_{\pm}\}$, 
among which those with particularly neat expressions are 
\beq
\widehat{\cal Y} = -2i \frac{\pa}{\pa\psi} 
\mbox{ } \mbox{ } , 
\mbox{ }  \mbox{ } 
\widehat{\cal I}_3 = -i\frac{\pa}{\pa\phi}  
\mbox{ } \mbox{ } ,   
\eeq
\beq
i\widehat{\cal I}_1 = - \mbox{sin}\,\phi\frac{\pa}{\pa \beta} + \frac{\mbox{cos}\,\phi}{\mbox{sin}\,\beta}
\left\{
\frac{\pa}{\pa\psi} - \mbox{cos}\,\beta\frac{\pa}{\pa\phi}
\right\} 
\mbox{ } \mbox{ } , \mbox{ } 
i\widehat{\cal I}_2 = \mbox{cos}\,\phi\frac{\pa}{\pa \beta} + \frac{\mbox{sin}\,\phi}{\mbox{sin}\,\beta}\left\{\frac{\pa}{\pa\psi} - 
\mbox{cos}\,\beta\frac{\pa}{\pa\phi}\right\} \mbox{ } \mbox{ } .  
\eeq
Finally,  
\beq
\hat{\cal I}^2 = -
\left\{
\frac{1}{\mbox{sin}\,\beta}  \frac{\pa}{\pa\beta}\mbox{sin}\,\beta\frac{\pa}{\pa\beta} + 
\frac{1}{\mbox{sin}^2\beta}
\left\{
\frac{\pa^2}{\pa\phi^2} - 
2\mbox{cos}\,\beta\frac{\pa}{\pa\psi}\frac{\pa}{\pa\phi} + \frac{\pa^2}{\pa\psi^2}
\right\}
\right\} \mbox{ } . 
\eeq
This is given this in the operator-ordering that is relevant to this paper's time-independent Schr\"{o}dinger equation to be in terms of the Laplacian (see the next Section).
These can then be paired with the Gibbons--Pope type coordinate expressions for the shape quantities $s^{\Gamma}$ of Sec I.16 so as to demonstrate closure and evaluate the commutators.  

\mbox{ } 

\noindent The kinematical quantization of GR-as-geometrodynamics itself involves 
${\cal C}^{\infty}(\bupSigma, M(3, \mathbb{R})) \, \mbox{\textcircled{S}} \,{\cal C}^{\infty}(\bupSigma, \mbox{GL}^+(3, \mathbb{R}))$ \cite{I84}.  
Here the latter factor is closely associated with the mathematical identity of Riem($\bupSigma$), GL stands for `general linear' and $M(3, \mathbb{R})$ are real 3 $\times$ 3 matrices.  

\mbox{ } 

\noindent \underline{Key 19} One can furthermore consider the above choice of kinematical quantization as also a selection of relational beables. 
I.e. a subset \{$K_{\sfA}$\} of the classical \K beables $K = F[\chi, \beta, \phi, \psi, p_{\chi}, p_{\beta}, p_{\phi}, p_{\psi}]$ 
then promoted to the quantum level \{$\widehat{K}_{\sfA}$\} such that 

\noindent 0) They obey [$\widehat{K}, \widehat{\mbox{Lin}}_{\sfZ}$] = 0 which is trivially the case here by prior classical reduction.

\noindent 1) They cover all the relational information.  

\noindent 2) They obey suitable continuity conditions.

\noindent 3) They themselves algebraically close under the commutation relation.

\noindent 4) They are allowed some redundancy 
(meaning more relational functions than there are independent pieces of relational information -- not to be confused with including unphysical/gauge/non-relational information).

\noindent Note that 2) to 4) are e.g. already evident in using sin$\,\phi$ and cos\,$\phi$ for the circle/3-stop metroland.

\noindent Then a candidate for the algebra of \K beables for quadrilateralland is that it is the same as the kinematical quantization algebra of the 
 $\widehat{T}_{\Gamma}$ and $s^{\Gamma}$.

\section{Time-independent Schr\"{o}dinger equation (TISE) for quadrilateralland}\label{TISE}

Operator-ordering is more of a problem in Quantum Cosmology than in Atomic Physics due to paucity of observations. 
Can theory alone determine operator ordering?

DeWitt \cite{DeWitt57} considered elevating the classical coordinatization-independence of configuration space to additionally hold at the quantum level. 
This suggests that the classical kinetic term $\fN^{\sfA\sfB}(Q^{\sfC})P^{\sK}_{\sfA}P^{\sK}_{\sfB}$ be promoted to the quantum-level {\sl Laplacian operator ordering} 
\beq
\triangle := 
\frac{1}{\sqrt{M(Q^{\sfC})}} \frac{\nabla}{\nabla {Q}^{\sfA}}
\left\{
\sqrt{{M}(Q^{\sfC})}{\fN}^{\sfA\sfB}(Q^{\sfC})\frac{\nabla}{\nabla Q^{\sfB}}
\right\} 
\eeq
(this is also advocated in e.g. \cite{K73etc}).
Moreover, this is not a unique implementation of DeWitt's criterion since one can include a Ricci scalar curvature term so as to have, for any $\xi \in \mathbb{R}$, \footnote{Even then, 
an underlying simplicity here is that the above is the extent of the ambiguity only if one  excludes more complicated curvature scalars. 
E.g. one excludes these by stipulating  no higher-order derivatives nor higher-degree polynomials in the derivatives.}
the {\sl $\xi$-operator ordering}
\beq
\triangle^{\xi} := \triangle - \xi\,\mbox{Ric}(Q; M]  \mbox{ } .  
\eeq 
\noindent Among these $\xi$-orderings, there is \cite{Wald} a unique configuration space dimension $q$-dependent ($q > 1$) {\sl conformally-invariant} choice 
of operator-ordering (\cite{Magic, ConfOrder}), 
\beq
\triangle^{\sc} := \triangle - \xi^{\sc} \mbox{Ric}(Q; M] := \triangle - \frac{q - 2}{4\{q - 1\}}\mbox{Ric}(Q; M] \mbox{ } . 
\eeq 
This furthermore requires that $\Psi$ itself transforms in general tensorially under conformal transformations \cite{Wald}, 
\beq
\Psi \longrightarrow \widetilde{\Psi} = \Omega^{\{2 - q\}/2}\Psi \mbox{ } .
\label{PPCTwave}
\eeq  
\noindent The TISE following from the above family of orderings is then ($\fE_{\sU\sn\si}$ denotes the total energy of the model universe) 
\beq
{\cal H} \Psi = \fE_{\sU\sn\si}\Psi \Rightarrow \triangle^{\sc}\Psi = 2\{\fV - \fE_{\sU\sn\si}\}\Psi/\hbar^2 \mbox{ } .  
\label{DanSci}
\eeq
\noindent What is the underlying conformal invariance in question? 
[E.g. it does {\sl not} act on space itself.]

\mbox{ } 

\noindent {\sl Misner's identification} \cite{Magic} is that it is the underlying conformal covariance of the Hamiltonian constraint under scaling transformations. 
\beq
{\cal H} = 0 \longrightarrow \widetilde{\cal H} := \Omega^{-2}{\cal H} = 0 \mbox{ } .
\eeq
This can be generalized to conformal covariance of other quadratic constraints such as the energy constraint ${\cal E}$ or the r-formulation counterpart ${\cal E}^{\sr}$.  
\noindent \underline{Key 20} {\sl E.A.'s identification} \cite{Banal}, on the other hand, goes one level deeper to the consideration of actions.  
It then so happens that it is the conformal invariance 
\beq
\d \fs^2 \rightarrow \d\widetilde{\fs}^2 = \Omega^2\d \fs^2 \mbox{ } , \mbox{ } \mbox{ } 
 \fE - \fV \rightarrow \{\widetilde{\fE}_{\sU\sn\si}  - \widetilde{\fV}\} = \{\fE_{\sU\sn\si} - \fV\}/\Omega^2  \mbox{ }   
\label{PPCT}
\eeq
of relational product actions [c.f. (I.1, 3, 4, 12) for examples]. 
I.e. it is a conformal invariance of the kinetic arc element $\d\fs$ alongside a compensatory conformal invariance in the potential factor $\fW$. 
This reflects that the combination actually present in the action, $\d\widetilde{\fs}$, is not physically meaningfully factorizable.  
One then recovers Misner's conformal covariance for the purely-quadratic constraint one's theory possesses as a primary constraint due to its relational product form. 

\mbox{ }  

\noindent For the physical quantities to be invariant, the inner product in this convention is to have a weight function $\omega$ scaling as (see e.g. \cite{08II})
\beq
\omega \longrightarrow \widetilde{\omega} = \Omega^{-2}\omega .  
\label{PPCTweight}
\eeq 
Thus   
\beq
\int \widetilde{\mathbb{D}Q \, \Psi_1^*\Psi_2 \omega} = 
\int \mathbb{D}Q \Omega^q  \, \Psi_1^*\Psi_2 \Omega^{2\{2 - q\}/2} \omega \Omega^{-2} = 
\int \mathbb{D}Q \, \Psi_1^*\Psi_2 \omega \mbox{ } .  
\eeq

\mbox{ }

\noindent Example 1) For triangleland and 4-stop metroland, q = 2 so $\xi^{\sc} = 0$ and conformal ordering = Laplacian ordering.

\noindent Example 2) For the general $\mathbb{CP}^{n - 1}$, $q = 2\{n - 1\}$ and Ric = $4n\{n - 1\}$. 
Thus $\triangle^{\sc}_{\mathbb{CP}^{n - 1}} = \triangle_{\mathbb{CP}^{n - 1}} - 2n\{n - 1\}\{n - 2\}/\{2n - 3\}$.
In particular, then, for quadrilateralland $\triangle^{\sc}_{\mathbb{CP}^{2}} = \triangle_{\mathbb{CP}^{2}} - 4$.

\mbox{ } 

\noindent The corresponding TISE's are then as follows.

\noindent For 4-stop metroland in terms of $\widehat{\cal D}_{\sT\so\st}$,   
\beq
\widehat{\cal D}_{\sT\so\st} \Psi = \triangle_{\mathbb{S}^{2}} \Psi = 2\{\fV - \fE\} \Psi/\hbar^2 \mbox{ } .  
\eeq
For $N$-a-gonland in general terms, 
\beq
\triangle_{\mathbb{CP}^{n - 1}}\Psi - 2n\{n - 1\}\{n - 2\}\Psi/\{2n - 3\} = 2\{\fV - \fE\}\Psi/\hbar^2 \mbox{ } .  
\eeq
Then specializing and further specifying for the triangle in terms of $\widehat{\cal S}_{\sT\so\st}$, 
\beq
\widehat{\cal S}_{\sT\so\st}\Psi = \triangle_{\mathbb{S}^2}\Psi = \{\fV - \fE\}\Psi/2\hbar^2 \mbox{ } ,
\eeq
whilst doing so for the quadrilateral in terms now of two distinct `felt charges' associated with ${\cal I}_{\sT\so\st}$ and ${\cal Y}$ gives 
\beq
- \frac{1}{\mbox{sin}^3\chi\,\mbox{cos}\,\chi}\frac{\pa}{\pa\chi}
\left\{
\mbox{sin}^3\chi\,\mbox{cos}\,\chi 
\frac{\pa \Psi}{\pa\chi}\right\} - \frac{\widehat{\cal I}_{\sT\so\st}\Psi}{\mbox{sin}^2\chi} - \frac{\widehat{\cal Y}^2\Psi}{\mbox{cos}^2\chi} +
\frac{2n\{n - 1\}\{n - 2\}}{2n - 3}\Psi = \frac{2\{E_{\sU\sn\si} - V\}}{\hbar^2}\Psi \mbox{ } .
\eeq
The GR-as-geometrodynamics counterpart of this is the Wheeler--DeWitt equation 
\beq
\widehat{\cal H}\Psi := - \hbar^2`\left\{\frac{1}{\sqrt{\fM}}  \frac{\delta}{\delta \mh^{\mu\nu}}
\left\{
\sqrt{\fM}\mN^{\mu\nu\rho\sigma}  \frac{\delta\Psi}{\delta \mh^{\rho\sigma}}
\right\} 
- \frac{\mbox{Ric}_{\sfM}(x^{\mu}; h_{\mu\nu}]}{4}\right\}\mbox{'}\Psi - \sqrt{\mh}\mbox{${\cal R}\mi\mc$}(x^{\mu}; h_{\mu\nu}]\Psi + 2\sqrt{\mh}\Lambda\Psi = 0  \label{WDE2} \mbox{ } ,     
\eeq
coupled to a QM momentum constraint in general. 
` ' here indicates regularization, well-definedness and operator-ordering issues.
The minisuperspace version has partial rather than functional derivatives, a dimension-dependent operator-ordering coefficient for Ric$_{\sfM}$ rather than the infinite-dimensional limit 
and no QM momentum constraint.

\section{Quadrilateralland QM separates in Gibbons--Pope type coordinates}\label{sep}

The hydrogen atom's TISE separates in both spherical and parabolic coordinates \cite{LLQM} and that of the isotropic HO in both Cartesian and spherical coordinates \cite{Messiah}.
However, it is generally regarded as quite good fortune to be able to separate a quantum problem at all. 
How do RPM's fare?  

\noindent The free and isotropic-HO type problems, $N$-stop metroland \cite{AF, ScaleQM} and triangleland \cite{08II, +tri, 08III} are separable in (ultra)spherical coordinates. 
\noindent $N$-stop metroland is also separable for these and the diagonal anisotropic HO type problem in the Cartesian coordinates that physically represent the 
relative Jacobi inter-particle cluster separations \cite{06II, FileR}.  
\noindent Scaled triangleland \cite{08III} is also separable for the above potentials and for the diagonal anisotropic HO in parabolic coordinates, 
that here physically signify a split into subsystems (`base and median').
For the quadrilateral, we {\sl have} specifically considered \cite{MacFarlane, QuadI} the Gibbons--Pope type coordinates as best-possible analogues of the (ultra)spherical coordinates.   
Indeed these do not disappoint when it comes to separability of the free (or isotropic-HO type) TISE: \underline{Key 21}.  
\noindent Quadrilateralland separates in Gibbons--Pope type coordinates.  
These coordinates are $SU(2) \times U(1)$-adapted, and that part separates out as a package [using $\Psi(\chi,\beta,\phi,\psi) = A(\psi, \phi, \beta)R(\chi)$].

\mbox{ } 

\noindent The free problems for pure-shape 4-stop metroland and triangleland then form the even more standard package solved by the spherical harmonics.  
(These are themselves separable into an SHM part and an associated Legendre part.) 
Moreover, pinning physical interpretation on these, they come as, firstly, 
\beq
\widehat{{\cal D}}_{\sT\so\st} \Psi(\theta, \phi) = \mD\{\mD + 1\} \Psi(\theta, \phi) \mbox{ } \mbox{ (pure-shape 4-stop metroland)} 
\eeq
for D total relative dilational momentum quantum number, with second quantum number d the relative dilational momentum between the base and median subsystems, 
featuring as the eigenvalue in ${\cal D}_3\Psi = \d \Psi$.  
Secondly, 
\beq
\widehat{{\cal S}}_{\sT\so\st} \Psi(\Theta, \Phi) = \mS\{\mS + 1\} \Psi(\Theta, \Phi) \mbox{ } \mbox{ (pure-shape triangleland)} 
\eeq
for S the total shape momentum quantum number: mixed relative angular and relative dilational momentum \cite{+tri, FileR}).  
Here, the second quantum number either is j the pure relative angular momentum component of the base relative to the median  
in the DES basis or an also mixed relative angular momentum and relative dilational momentum shape quantum number s in the EDS basis. 
This features in the eigenvalue problem ${\cal S}_3\Psi = \mj\Psi$ or $\ms\Psi$ (which depends on the meaning of the principal `3' axis in each coordinate basis).  
These are clearly analogues of the angular momentum eigenvalue equations for the rotor and atom.  
For quadrilateralland,
\beq
\widehat{{\cal I}}_{\sT\so\st} A(\psi, \beta, \phi ) = \mI\{\mI + 1\} A(\psi, \beta, \phi) \mbox{ } : 
\label{A-part}
\eeq
the `angular' part of the `angular to `radial' split, while the `radial' coordinate $\chi$ obeys, for our conformally-ordered free problem,  
\beq
- \frac{1}{\mbox{sin}^3\chi\,\mbox{cos}\,\chi}\frac{\d}{\d\chi}
\left\{
\mbox{sin}^3\chi\,\mbox{cos}\,\chi \frac{\d R(\chi)}{\d\chi}
\right\} 
+ \frac{4\mI\{\mI + 1\}R(\chi)}{\mbox{sin}^2\chi}  + \frac{\mY^2R(\chi)}{\mbox{cos}^2\chi} + \frac{2n\{n - 1\}\{n - 2\}}{2n - 3}R(\chi) = \frac{2E_{\sU\sn\si}}{\hbar^2} R(\chi) 
\mbox{ } .  
\eeq
Now, (\ref{A-part}) -- analogous to the SHM part of the spherical harmonics equation -- is a somewhat more complicated but still standard equation 
solved by the Wigner-D functions (see Appendix C).  
The quantum number I is interpreted as `total isospin', i.e. in H-coordinates (Sec I.10), the total angular momentum of the posts, and in K-coordinates (Sec I.10) 
the picked-out `axe blade' triangle subsystem.
The $\mI_3$ and $\mY$ quantum numbers are most candidly as the eigenvalues in
\beq
\widehat{\cal I}_3 \mD(\psi, \beta, \phi) = \mI_3        \mD(\psi, \beta, \phi) \mbox{ } \mbox{ and } \mbox{ } \mbox{ }
\widehat{\cal Y}   \mD(\psi, \beta, \phi) = \frac{\mY}{2}\mD(\psi, \beta, \phi) \mbox{ } .   
\eeq
$\mI_{3}$ is thus interpreted as `3-component of isospin'. 
I.e. in H-coordinates it is the counter-rotation of the two posts relative to the crossbar. 
In K-coordinates it is the co-rotation of the face of the `axe blade' and the handle relative to the depth of the blade. 
(See Fig I.10 for nomenclature.)  
Also, Y is thus interpreted as the `hypercharge' i.e. `extra angular charge' quantum number. 
In H-coordinates, it is the co-rotation of the two posts relative to the crossbar. 
In K-coordinates, it is the counter-rotation of the face of the `axe blade' and the handle relative to the depth of the blade.  
\noindent The ranges for the quantum numbers are 2$\mI \, \in \, \mathbb{N}_0$ and $2\mI_3, \mY \in \mathbb{Z}$  with $-\mI \leq \mI_3, \mY/2 \leq \mI$.   
Finally, the `radial' $\chi$-equation -- analogous to the associated Legendre equation part of the spherical harmonics equation -- can be shown \cite{MacFarlane} to map to the 
hypergeometric equation. 
Thus it is solved by Jacobi polynomials (within which the Gegenbauer and Legendre polynomials are nested, see Appendix B).  
Wigner D-functions are a combination of elementary trig functions and, once again, Jacobi polynomials (see Appendix C).  
They arise also in the elementary study of {\sl finite} rotations in QM.

\section{Free problem's general solution and quadrilateralland interpretation}\label{free-solve}

\noindent For useful contrast, specializing (\ref{A1}) and (\ref{A2}) to the $\mN = N - 2 = 1$ of pure-shape triangleland,
\beq
{\cal E}(k, 1) = 4 k\{k + 1\} \mbox{ } , \mbox{ } \mbox{ } k \, \in \mbox{ } \mathbb{N}_0 \mbox{ } .  
\eeq
This is very familiar as a proportionality when $k$ is denoted by l or J in the rigid rotor (though for triangleland itself the quantum number is denoted by S for `shape').  
The degeneracies are  
\beq
{\cal D}(k, 1) = 2k + 1
\eeq
which is also very familiar [$SU(2)$ multiplets].
Thus this model has 
1  ground state `$s$-orbital', 
3  first excited state `$p$-orbitals' and 
5  second excited state `$d$-orbitals'
using the spectoscopic notation familiar from Atomic Physics.  
For the triangle these are mathematically the same as for the atom but have the distinct physical interpretation provided in \cite{+tri, FileR}.  
Free 4-stop metroland has the above eigenvalues and multiplicities too \cite{AF, FileR}, only now the quantum number involved is denoted by D for `dilational'.
Free $N$-stop metroland \cite{AF, ScaleQM} exhibits the reasonably well known `rotor in $N - 1$ dimensions' pattern.  
This is also the mathematical basis of the simplest `$N - 1$-dimensional analogue of the periodic table' with $N - 1$ `$p$-orbitals' and $N\{N - 1\}/2 - 1$ `$d$-orbitals'.

\noindent Next, specializing (\ref{A1}) and (\ref{A2}) to the $\mN = N - 2 = 2$ of $\mathbb{CP}^2$ \cite{W82b, MacFarlane}),  
\beq
{\cal E}(k, 2) = 4k\{k + 2\}
\label{naive-CP2-En}
\eeq
with degeneracies
\beq
{\cal D}(k, 2) = \{k + 1\}^3 \mbox{ } .  
\eeq
This has 
1  ground state               `$s$-orbital', 
8  first excited state      `$p$-orbitals' and 
27 second excited state     `$d$-orbitals'. 
In this case, $k := \mI + \mY/2 + \mn$, indeed motivated by being the sole functional dependence on the quantum numbers in the expression for the energy. 
Here $\mn \, \in \, \mathbb{N}_0$ the degree in $\mbox{cos}\,2\chi$ of the Jacobi polynomial in the corresponding expressions for the solutions at the energy level in question.  
This has some parallels with the principal quantum number of the atom and with its counterpart for the isotropic HO; these are all `radial node counting' quantum numbers. 
The role of `radius' in the current pure-shape problem is played by the $\chi$ coordinate. 
[I.e it is radial in contradistinction to the Euler angles' $SU(2)$-angularness. 
This is all to be taken within the context of the geometrically of configuration space rather than of the physics in space.]  

\noindent However, the above eigenvalues for Berger et al's and MacFarlane's treatments of $\mathbb{CP}^2$ only involve Laplacian operator ordering  and no factor of $\hbar^2/2$.  
Moreover, Ric($Q; M$] is but constant here, so for quadrilateralland one can just take on MacFarlane's equation for a shifted energy as per above:    
\beq
\fE_{\sU\sn\si} = \hbar^2{\cal E}/{2} - \hbar^2 2n\{n - 1\}\{n - 2\}/\{2n - 3\} \mbox{ } . 
\label{NagonlandshiftE}
\eeq
The second term here is $- 4\hbar^2$ for the quadrilateral. 
Thus 
\beq
\fE_{\sU\sn\si} = 2\hbar^2\{k + 1\}^2 \mbox{ } , \mbox{ } \mbox{ } k \in \mathbb{N}_0 \mbox{ } .  
\eeq
\noindent Next, capitalizing on the split and identification at the end of Sec \ref{sep}, the general wavefunctions for quadrilateralland are 
$$
\Psi_{\sn\,\sI\,\sI_3\,\sY}(\chi, \beta, \phi,\psi) = \frac{1}{\sqrt{\pi}} 
\left\{
\frac{\{2\{\mn + \mI + 1\} + |\mY|\} \Gamma(\mn + 2\mI + |\mY| + 2)\mn!}{\Gamma(\mn + 2\mI + 2)\Gamma(\mn + |\mY| + 1)}
\frac{\Gamma(\mI + \mY/2 + 1)\Gamma(\mI - \mY/2 + 1)  }{\Gamma(\mI + \mI_3 + 1)\Gamma(\mI - \mI_3 + 1)}
\right\}^{1/2} \times
$$
\beq
\mbox{sin}^{2\sI}\chi \mbox{cos}^{|\sY|}\chi\mP^{(2\sI + 1,|\sY|)}_{\sn}(\mbox{cos}\,2\chi)\mbox{exp}(i\,\mY\,\psi/2)\md^{\sI}_{{\sY}/{2}, \sI_3}(\beta)\mbox{exp}(i\,\mI_3\phi) 
\mbox{} \mbox{ } .   
\eeq
\noindent \underline{Key 22} This is obtained by interpreting MacFarlane's study \cite{MacFarlane} of the QM of $\mathbb{CP}^2$ in quadrilateralland terms. 
Additionally the normalization coefficients (that he omitted) are required since we need them for our subsequent calculations.  
\noindent The mod bars come from the need for Re\,$\alpha$, Re\,$\beta > -1$ in the theory of the Jacobi polynomials (Appendix B).  
In this way we offer a minor correction of eq. (85) of MacFarlane \cite{MacFarlane}.  
The rest of the Y's present do not need mod bars by symmetry and the range of definition of the Wigner-d and exponential functions. 
This is likewise for all of the $\mI_3$'s.  

\mbox{ } 

\noindent For contrast, for the pure-shape versions of 4-stop metroland and triangleland are, respectively, 
\beq
\Psi_{\sD\,\sd} \propto Y_{\sD\,\sd}(\Theta,\Phi) \propto \mP^{|\sd|}_{\sD}(\mbox{cos}\,\theta)\mbox{exp}(i\,\md\,\phi) \mbox{ } ,
\eeq
\beq
\Psi_{\sS\,\sj} \propto Y_{\sS\,\sj}(\Theta,\Phi) \propto \mP^{|\sj|}_{\sS}(\mbox{cos}\,\Theta)\mbox{exp}(i\,\mj\,\Phi) \mbox{ }  .  
\eeq
Here $\mP$ now the associated Legendre functions.  
Thus we have a pure-ratio factor Rat($\Theta$) and a pure-relative-angle factor Ang($\Phi$) in the triangleland case.\footnote{Pure-ratio 
here means a purely non-angular ratio, angles themselves of course being expressible in terms of certain other ratios. 
This all refers to the status of these coordinates interpretations {\sl in space}. 
E.g. $\theta$ is an angle in configuration space, but is physically a function of a purely non-angular ratio in space itself.}
%
This feature is indeed repeated in quadrilateralland: a pure-ratio factor Rat($\chi, \beta$) and a pure-relative-angle factor Ang$(\phi, \psi$).  
Note that while the Wigner D-function is a useful {\sl solving} package, it is not aligned with this useful {\sl interpretational} split. 
However, since it is itself separable into functions of all three variables, there is no problem in refactorizing the wavefunction.

Thus both 4-stop metroland and triangleland free pure-shape problems have rigid-rotor mathematics. 
On the other hand, the 4-stop metroland case is physically a relative dilatator and the triangleland case is a mixed relative rotor relative dilatator.  
This is probably best called a `rationator'. 
Quadrilateralland is again a rationator, and one no longer constrained to obey $SU(2)$ mathematics so that it is not the same as a rigid rotor.  
%
%
The symmetrical, alias Lagrange, spinning top with $I_1 = I_2 \neq I_3$, 
does itself involve at the quantum level eigenfunctions based on the Jacobi Polynomials \cite{Edmonds}. 
However, these are mathematically distinct from the current paper's in greater detail.  

\mbox{ } 

\noindent Parallelling further 4-stop metroland and triangleland calculations in \cite{AF, +tri}, e.g. 
\beq
\mbox{sin}^{2\sI}\chi \mbox{cos}^{|\sY|}\chi\mP^{(2\sI + 1, |\sY|)}_{\sn}(\mbox{cos}\,2\chi)\mbox{cos}(\mY\,\psi/2)\md^{\sI}_{{\sY}/{2}, \sI_3}(\beta)\mbox{cos}(\mI_3\phi) 
\mbox{} \mbox{ } 
\eeq
can be recast as 
\beq
\{1 -  \mn_3^2\}^{\sI}\chi \mn_3^{|\sY|}  
\mP^{(2\sI + 1, |\sY|)}_{\sn}(2\mn_3^2 - 1)
\mbox{\Huge$\{$}
\stackrel{\mbox{\normalsize $\mT_{\sY/2}(\mn_2\cdot\mn_3)\mT_{\sY/2}(\mn_1\cdot\mn_3)$}}{- {\cal T}_{\sY/2}(\mn_2\cdot\mn_3){\cal T}_{\sY/2}(\mn_1\cdot\mn_3)}
\mbox{\Huge$\}$}
\md^{\sI}_{{\sY}/{2}, \sI_3}(\mbox{arctan}(\mn_2/\mn_1))
\mbox{\Huge$\{$}
\stackrel{\mbox{\normalsize $\mT_{\sI_3}(\mn_2\cdot\mn_3)\mT_{\sI_3}(\mn_1\cdot\mn_3)$}}{+ {\cal T}_{\sI_3}(\mn_2\cdot\mn_3){\cal T}_{\sI_3}(\mn_1\cdot\mn_3)}
\mbox{\Huge$\}$}
\mbox{ } \mbox{ }    
\eeq
for ${\cal T}_{\sp}(X) := \sqrt{1 - T_{\sp}(X)^2}$
and $T_{\sp}(X)$ the Tchebychev polynomial of the first kind of degree p in $X$ (see Appendix B).

\section{Visualization and discussion of first few wavefunctions}\label{vis}

\noindent This account parallels \cite{AF} and \cite{08II, +tri} of 4-stop metroland and triangleland respectively. 
There, immediately visualizable 2-$d$ tessellations were available.  
However, now for quadrilateralland, we are at some disadvantage.  
Though at least some of the simpler wavefunctions can be viewed without loss with some dimensions suppressed.  

\mbox{ } 

\noindent The ground state is
\beq
s = \Psi_{0\,0\,0\,0} = \sqrt{2}/\pi \mbox{ } , \mbox{ } \mbox{ const } .  
\eeq
Thus it favours no particular regions or directions,\footnote{This is an ubiquitous feature -- c.f. 
the ground state that is constant over the (k-)sphere for standard rotors and the corresponding $N$-stop metroland and triangleland problems.}  
here meaning types of quadrilateral.
 
\noindent 1) The first harmonics are the octet 

\noindent
\beq
p_{\Lambda} := \Psi_{1\,0\,0\,0}   = \{2/\pi\}\{1 - 3\,\mbox{cos}^2\chi\} \mbox{ } , \mbox{ } \mbox{ }
\eeq
\beq
p_{\Sigma_0} := \Psi_{0\,1\,0\,0}  = \{2/\pi\}\mbox{sin}^2\chi\,\mbox{cos}\,\beta \mbox{ } , \mbox{ } \mbox{ }
\eeq
\beq
p_{\Sigma_-} := \Psi_{0\,1\,-1\,0} =  \{\sqrt{2}/\pi\}\mbox{sin}^2\chi\,\mbox{sin}\,\beta\,\mbox{exp}(-i\,\phi)   \mbox{ } , \mbox{ } \mbox{ } 
p_{\Sigma_+} := \Psi_{0\,1\,1\,0}  = -\{\sqrt{2}/\pi\}\mbox{sin}^2\chi\,\mbox{sin}\,\beta\,\mbox{exp}( i\,\phi)  \mbox{ } , \mbox{ } \mbox{ }
\eeq
$$
p_{\sn} :=     \Psi_{0\,\half\,-\half\, 1} = \{\sqrt{3}/\pi\}\mbox{sin}\,2\chi\, \mbox{cos$\frac{\beta}{2}$} \, \mbox{exp}(-i\,\phi/2) \mbox{exp}( i\,\psi/2) 
\mbox{ } , \mbox{ } \mbox{ } 
$$
$$
p_{\sp} :=     \Psi_{0\,\half\, \half\, 1} = \{\sqrt{3}/\pi\}\mbox{sin}\,2\chi\, \mbox{cos$\frac{\beta}{2}$} \, \mbox{exp}( i\,\phi/2) \mbox{exp}( i\,\psi/2) 
\mbox{ } , \mbox{ } \mbox{ } 
$$
$$
p_{\Xi_-} := \Psi_{0\,\half\,-\half\,-1} = \{\sqrt{3}/\pi\}\mbox{sin}\,2\chi\, \mbox{sin$\frac{\beta}{2}$} \, \mbox{exp}(-i\,\phi/2) \mbox{exp}(-i\,\psi/2) 
\mbox{ } , \mbox{ } \mbox{ } 
$$
\beq
p_{\Xi_0} := \Psi_{0\,\half\, \half\,-1} = \{\sqrt{3}/\pi\}\mbox{sin}\,2\chi\, \mbox{sin$\frac{\beta}{2}$} \, \mbox{exp}( i\,\phi/2) \mbox{exp}(-i\,\psi/2) 
\mbox{ } . 
\eeq
%
{           \begin{figure}[ht]
\centering
\includegraphics[width=0.35\textwidth]{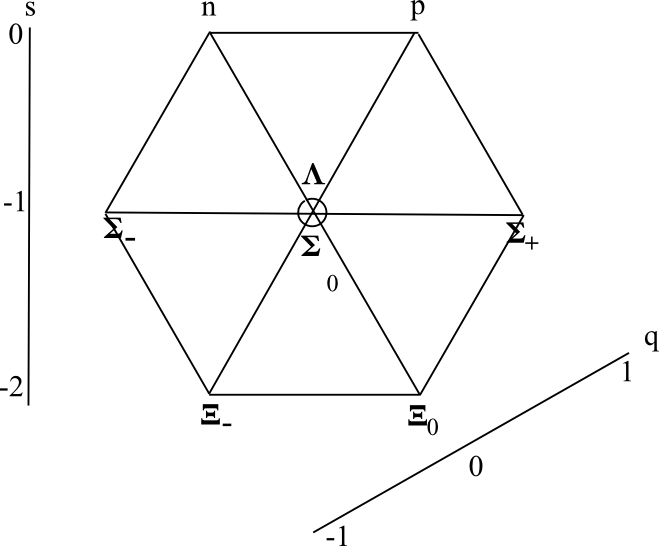}
\caption[Text der im Bilderverzeichnis auftaucht]{  \footnotesize{Gell-Mann's eightfold way multiplet familiar from Particle Physics.  
s is strangeness and q is charge.
p and n are the proton and the neutron.}        } 
\label{Eightfold} \end{figure}         } 

%
\noindent Note how quadrilateralland's orbitals, unlike those of the atom, its $SO(d)$ generalization in dimension $d$, $N$-stop metroland \cite{AF} and triangleland \cite{08II, +tri}, 
are no longer simply related to an obvious Cartesian space's axes. 
Thus they require a different kind of nomenclature for their labels.  
\noindent We still prefer sine and cosine combinations to $\pm$'s, 0/--'s and p/n's, in parallel with the preferred representations of the atomic orbitals.
See Fig 1 for the labelling nomenclature used in (34--38).  

Alternative specifically-quadrilateralland names for these orbitals are given in Figure \ref{Choppy}. 
This sketches each's p.d.f. over our complex-projective chopping board generalization of Kendall's spherical blackboard in the case for which the Gibbons--Pope type coordinates are 
adapted about a Jacobi K-tree.  
%
{           \begin{figure}[ht]
\centering
\includegraphics[width=0.85\textwidth]{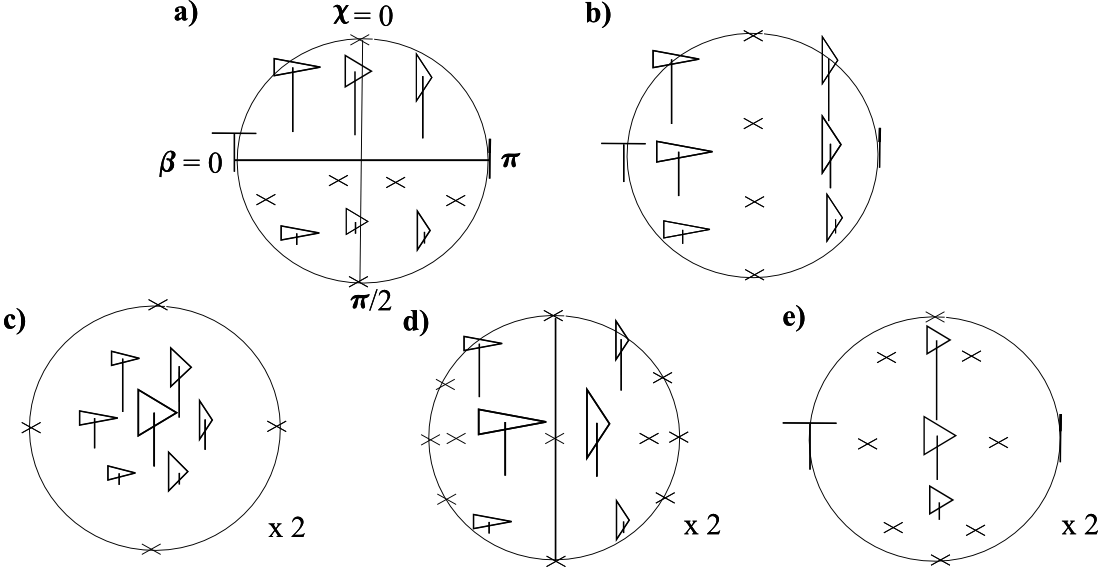}
\caption[Text der im Bilderverzeichnis auftaucht]{  \footnotesize{Let us adopt Fig I.10's complex-projective chopping board geometrical representation and its associated nomenclature. 
\noindent Then a) $p_{\Lambda}$'s is `long-and-short handled', though it slightly favours long over short by its nodal curve lying slightly into the short hemi-$\mathbb{CP}^2$.
\noindent b) $p_{\Sigma_0}$'s is `pick-or-executioner bladed'.  
\noindent c) $p_{\Sigma_{\pm}}$ and sine/cosine combinations thereof are `regular'.  
Further distinctions between these and within each of the below depend on additional relative-angle information.  
\noindent d) $p_{\Sigma_{0\,\,\so\sr\,\,-}}$  and sine/cosine combinations thereof are `moderately pick-or-executioner'
\noindent e) $p_{\sn}$ and $p_{\sp}$ and sine/cosine combinations are `all bar moderately pick-or-executioner'. 
\noindent x2 denotes that there are 2 distinct states with that pure-ratio profile.  
\noindent See Chapters 14 and 15 of \cite{FileR} for 3- and 4-stop metroland and triangleland spherical blackboard equivalents of this Figure.}        } 
\label{Choppy} \end{figure}         } 

By this stage we can see that the $p$'s can be identified octet of shape variables (modulo normalization).  
I.e. we have the `theorem' 
\beq
p_{\Gamma} \propto s^{\Gamma} \mbox{ } . 
\eeq
This is modulo taking sine and cosine combinations of the exponentials in the expressions for n and p, for $\Sigma_{\pm}$ and for $\Xi_{\pm}$ and re-ordering the basis to 
$(\Sigma_{\sc}, \Sigma_{\sss}, \Sigma_0,  (n, p)_{\sc}, (n, p)_{\sss}, \Xi_{\sc}, \Xi_{\sss}, \Lambda)$.
It is along the same lines (`naming polynomial' \cite{AF}) as how using Cartesians causes atomic orbitals to {\sl be} the functions they are named after.
(This is subject to minor conventions such as that $d_{z^2}$ is a contraction of $d_{3z^2 - 1}$.)  
It is clear then that the deep-seated way of labelling orbitals for QM based on $SU(n)$ is in terms of homogeneous polynomials rather than Cartesian axes. 
(Though both coincide for the well-known $SO(3)$--$SU(2)$ case.)
Thus we have obtained a distinct, specifically quadrilateralland-based nomenclature for the orbitals in terms of the shape quantities.
All in all, the 8 Gell-Mann quadratic forms are natural successors of triangleland's 3 Dragt quantities (that E.A. also termed Pauli quadratic forms). 
I.e. both sets are shape quantities, good for kinematical quantization and then a natural choice of labels for the quantum states for the corresponding free problem.

With respect to the $SU(2)$ privileged in this presentation, we have a singlet ($\Lambda$), a triplet (the $\Sigma$'s) and two doublets (p/n and the $\Xi$'s).  
\noindent The particular symmetries of the states are
$p_{\Lambda}$     is  $SU(2) \times U(1)$ symmetric, 
$p_{\Sigma_0}$    is  $U(1)  \times U(1)$ symmetric, the
$p_{\Sigma_{\pm}}$ are $U(1)$ symmetric and the rest have no continuous symmetries.

\section{Extension to scaled quadrilateralland QM (\underline{Key 23})}\label{scale}

Firstly, one gets a much closer match to GR quantum cosmology if one extends to the scaled C($\mathbb{CP}^2$) as in \cite{QuadI}.   
This is according to the following correspondences. 
I.e. i) configuration space radius $\rho$ to scale factor $a$.  
ii) Shape degrees of freedom to inhomogeneities.
iii) `Energy equation divided by moment of inertia' to `Friedmann equation post use of energy--momentum conservation equation' (see Chapter 5 of \cite{FileR} for details).  
The RPM potential can furthermore be chosen so that this analogue Friedmann equation parallels quantum cosmological scale dynamics.  
This has the further qualitative benefits that the associated small inhomogeneity mathematics is a lot more tractable for these models than in the actual Halliwell--Hawking scheme for GR 
itself.\footnote{The classical equations for this are in \cite{FileR}, at least for $\rho$, $\mZ$ coordinates.  
The C($\mathbb{CP}^2$) equations in terms of $\rho$ and Gibbons--Pope type coordinates may be new to us.  
We do not claim more than the radial conformal Killing vector, though it is a gap in our understanding as to whether this case furnishes more.  }

\subsection{Kinematical quantization}

\noindent{\bf Lemma}. Suppose one has a kinematical quantization algebra $\FC$ for a shape space. 
Then if the corresponding relational space has no `extra' symmetries, the kinematical quantization algebra of the corresponding relational space is 
${\FC} \, \mbox{\textcircled{S}} \, {\FA}\mathfrak{ff}$ for ${\FA}\mathfrak{ff}$ the `radial'/$\mathbb{R}_+$ problem's affine algebra.  

\mbox{ }

\noindent Example 1) Scaled $N$-stop metroland is exceptional due to possessing a number of extra symmetries.  
Moreover, this particular case's mathematics is, of course well-known by analogy with standard angular momentum.
Then the totality of the cone's symmetries that are not shape space symmetries are the translations.  
E.g. for 4-stop metroland,  one has $\fG_{\sc\sa\sn} \, \mbox{\textcircled{S}} \, \fV^*$ = $\{SO(3) \, \mbox{\textcircled{S}} \, \mathbb{R}^3\} \, \mbox{\textcircled{S}} \, \mathbb{R}^3$. 
This is mathematically a Heisenberg group. This has a second $SO(3)$-vector commutator and the standard commutation relation between the two conjugate vectors.

\noindent Example 2) Scaled triangleland is also exceptional, working mathematically just like the $n = 3$  
case of the preceding but physically the $\rho^i$ and $\pi_i$ are now, rather $\mbox{Dra}^{\Gamma}$ and $\Pi^{\sD\sr\sa}_{\Gamma}$.  

\noindent Example 3) Finally, for $N$-a-gonland, $N > 3$, by Sec \ref{KinQuant} and the Lemma, we take 
$\fG_{\sc\sa\sn} \, \mbox{\textcircled{S}} \, \fV^*$ = $SU(n) \, \mbox{\textcircled{S}} \,$  IHP($\mathbb{C}^n$, 2) $\, \mbox{\textcircled{S}} \,   {\FA}\mathfrak{ff}$.  
Here, the sole extra nontrivial commutation relation is the affine one, $[\rho, \pi] = i\hbar\rho$. 
[$\pi$ has to be represented as $-i\hbar\!\rho\!\pa_{\rho}$ in order to succeed in being self-adjoint.]

\subsection{Conformal ordering and the TISE}

\beq
\triangle^{\sc}_{\sC(\mathbb{CP}^{n - 1})} = \triangle_{\sC(\mathbb{CP}^{n - 1})}\Psi - {3\{2 n - 3\}}/{4\rho^2}  
\mbox{ }  
\mbox{ for } 
\mbox{ }  \mbox{ } 
\triangle_{\sC(\mathbb{CP}^{n - 1})} = \pa_{\rho}^2 + 2\{n - 1\}\rho^{-1} + \rho^{-2}\triangle_{\mathbb{CP}^{n - 1}} \mbox{ } . 
\eeq
Then the TISE is
\beq
-\{\pa_{\rho}^2 + 2\{n - 1\}\rho^{-1} + \rho^{-2}\{\triangle_{\mathbb{CP}^{n - 1}} - 3 \{2 n - 3\}/4\}\Psi = 2\{E_{\sU\sn\si} - V\}/\hbar^2 \mbox{ } .  
\eeq

\subsection{Scale--shape separation of the TISE}

Firstly note that the coning construction by which RPM's incorporate scale does not really care about the nature of the shape part, 
so the new split-out part comes out much the same as in \cite{ScaleQM}.
For $V = V(\rho)$ alone separability ensues [$\Psi = {\cal S}(\mbox{shape alone}){\cal R}(\rho)$].
Then 
\beq
\rho^2{\cal R}^{\prime\prime} + 2\{n - 1\}\rho{\cal R}^{\prime} + \{\{2\rho^2\{E_{\sU\sn\si} - V(\rho)\}/\hbar^2 -C\}{\cal R} = 0 \mbox{ } . 
\label{Nagon-chi}
\eeq
As a new feature from quadrilateralland upwards, the separated-out shape part gives a {\sl different} constant energy shifting, rather the conformal-ordered, pure-shape problem, 
\beq
\{\triangle_{\sC(\mathbb{CP}^{n - 1})} + C - 3\{2 n - 3\}/4\}S = 0 \mbox{ } , 
\eeq  
Comparing the first of these and the equation in Appendix A, we get that 
\beq
C = 4k\{k + n - 1\} + 3\{2 n - 3\}/4 =  4k\{k + 2\} + 9/4 \mbox{ }  \mbox{ for quadrilateralland } \mbox{ } . 
\eeq
$V = \kappa$ constant gives but an equation that maps to (Appendix D) the Bessel equation,
\beq
\rho^2{\cal R}^{\prime\prime} + 2\{n - 1\}\rho{\cal R}^{\prime} + \{2\{E_{\sU\sn\si} - \kappa\}\rho^2/\hbar^2 - 4k\{k + n - 1\} - 3\{2 n - 3\}/4 \}{\cal R} = 0 \mbox{ } ,
\eeq
whilst $V = A\rho^2$ gives but an an equation that maps to (Appendix D) the associated Laguerre equation,
\beq
\rho^2{\cal R}^{\prime\prime} + 2\{n - 1\}\rho{\cal R}^{\prime} + \{2\{E_{\sU\sn\si} - \kappa\}\rho^2/\hbar^2 - 2A\rho^4  -  4k\{k + n - 1\} - 3 \{2 n - 3\}/4 \}{\cal R} = 0 \mbox{ } . 
\eeq
This is similar to the radial equation for each of the atom and the isotropic HO.

\subsection{General solution}

Thus the solution to the first of these for general $N$-a-gonland is  
\beq
{\cal R} \propto \rho^{\{3n - 2\}/2}\mJ_{\pm\sqrt{4n^2 + nk + k^2 + k - 6n}}
\left( 
\sqrt{2\{E_{\sU\sn\si} - \kappa\}}\rho/\hbar 
\right)
= {\cal R} \propto \rho^{\{3n - 2\}/2}\mJ_{\pm\sqrt{2\{9 + 8k + 2k^2\}}}
\left( 
\sqrt{2\{E_{\sU\sn\si} - \kappa\}}\rho/\hbar \mbox{ }  
\right)
\eeq
with the second equality specializing to quadrilateralland.
This is simpler than the below, but has uncontained/un-normalizable character which is undesirable, so we mostly use the second example.

For that, the general-$N$ wavefunction is 
\beq
R_{\sn}(\rho) \propto \rho^{\{3 - 2n\}/2 + \Lambda_{n}}            \mL^{\Lambda_n}_{\sn}(\omega\rho^2/\hbar)                     \mbox{exp}(-\omega\rho^2/2) \mbox{ } .  
\eeq
Then the quadrilateralland case is just the $\Lambda_n$ to $\Lambda$ subcase of this.
Here, \noindent $E = \hbar\omega\{2\{\mn + \Lambda\} + 1\}$ for $\mn \in \mathbb{N}$, 

\noindent $\Lambda_n := \sqrt{\{2n - 3\}\{2n - 9/4\} + 16\mT\{\mT + n - 1\}}\} = \sqrt{45 + 64\mT\{ \mT + 2\}}/2$ for the quadrilateral, 
and $\omega := \sqrt{K} = \sqrt{2A}$.

\noindent These are roughly like the well-known radial profiles for atoms (themselves described by associated Laguerre functions) at least in terms of numbers of peaks and nodes.
One qualitative difference is that atoms and $N$-stop metroland have the outer peak of the ``2$s$" orbital p.d.f. much larger than the outer one.  
This corresponds to the most of the ``2$s$" orbital lying outside the ``1s" one, whereas this paper's $N$-a-gonland models have these two p.d.f. peaks of the same area as each 
other to within a few percent.

\section{A family of useful integrals (\underline{Key 24})}\label{Use-Int}

\noindent We consider integrals of the form 
\beq
\langle \psi_1\,|\, \widehat{O} \,|\,\psi_2\,\rangle \mbox{ } .
\eeq 
Subcases of these include the overlap integrals (for which the inserted operator $\widehat{O}$ = id) and the expectation values (for which $\psi_1 = \psi_2$).
Specific cases include $\langle \psi_1\,|\, \sigma^{\sn} \,|\,\psi_2\,\rangle$ (for powers of one's model's scale variable, $\sigma$), and 
$\langle \psi_1\,|\, \mbox{cos}\,\alpha \,|\,\psi_2\,\rangle$ or $\langle \psi_1\,|\, \mbox{cos}^2\alpha \,|\,\psi_2\,\rangle$ for $\alpha$ a non-scale (i.e. preshape) variable.

\noindent Three applications of these integrals are as follows.  

\noindent Application 1) expectation and spread of the scale and non-scale quantities in question.  

\noindent Application 2) Time-independent perturbation theory about e.g. free or HO-potential exact solutions.  
For quadrilateralland, however, it is not presently clear whether the extra indices that the Jacobi polynomials possess substantially complicate these calculations 
relative to those for the atom/triangleland by providing additional types of transition channels/selection rules.
Thus we do not yet know how to proceed to the particularly significant second-order case of this application. 
This is since this involves unequal quantum numbers on the 2 input wavefunctions as per the well-known general formula\footnote{Here, 
$\Xi$ and $\Omega$ are multi-indices running over all of the system's quantum numbers.}
\beq
E_{\Xi\,\Omega}^{(2)} = -\sum\mbox{}_{\mbox{}_{\mbox{\scriptsize $\Xi, \Omega$, $E_{\Xi} \neq E_{\Omega}$}}} 
      |\langle \Xi|\,{V}^{\prime}\,| \Omega \rangle|^2/
      \{ E_{\Xi} - E_{\Omega}    \} \mbox{ }   
\label{p2}
\eeq
for all that the simpler first-order formula 
\beq
E_{\Omega}^{(1)} = \langle \Omega\,|\,{V}^{\prime}\,|\, \Omega \rangle  \mbox{ } 
\label{p1}
\eeq
is under control.
We leave this point to a subsequent paper, noting that no such extra effects have for now been reported \cite{Norway} for the intermediate-difficulty 
(Gegenbauer polynomials, see Appendix B) problem of the Stark Effect in higher dimensions.  

\noindent Application 3) The really significant application for the current program is, moreover time-dependent perturbation theory on the space of shapes 
with respect to the emergent time provided by the scale in the scale--shape split of scaled RPM models.  
This is useful due to its analogy with the Semiclassical Approach to the PoT and Quantum Cosmology.

\noindent Note that both Applications 2) and 3) involve integrals of the specific form 
\beq
\langle \psi_1| V^{\prime}| \psi_2\rangle \mbox{ } , 
\eeq
for $V^{\prime}$ the perturbation part of the potential.  

\mbox{ }  

%
\noindent Atomic Example 1) From the angular factors of the integrals trivially cancelling and orthogonality and recurrence relation properties 
of Laguerre polynomials for the radial factors \cite{Messiah}, 
\beq
\langle\mn\,\ml\,\mm\,|\,r\,|\,\mn\,\ml\,\mm\rangle = \{3\mn^2 - \ml\{\ml + 1\}\}a_0/2 \mbox{ } \mbox{ and } \mbox{ } 
\Delta_{\sn\,\sll\,\sm}r = \sqrt{\{\mn^2\{\mn^2 + 2\} - \{\ml\{\ml + 1\}\}^2\}}a_0/2 \mbox{ } ,
\eeq
where $a_0$ is the Bohr radius of the atom.
One can then infer from this that 1) a minimal characteristic size is $3a_0/2$ for the ground state. 
%
%
%
2) The radius and its spread both become large for large quantum numbers; moreover, for these, Brown showed that the classical orbits are well-approximated \cite{Brown}.  

\noindent Atomic Example 2) $\langle\mn^{\prime}\,\ml^{\prime}\,\mm^{\prime}\,|\,\mbox{cos}\,\theta_{\sss\sp}\,|\,\mn\,\ml\,\mm\rangle$ and 
$\langle\mn^{\prime}\,\ml^{\prime}\,\mm^{\prime}\,|\,\mbox{cos}^2\,\theta_{\sss\sp}\,|\,\mn\,\ml\,\mm\rangle$ are 3-$Y$ integrals \cite{LLQM}.
(I.e. products of three spherical harmonics, $Y_{\sJ\sK}$, the radial parts of the integration now trivially cancelling.) 
Here `sp' denotes that the angles are taken in the the spatial sense that is common elsewhere than in this paper.   
Then the general case of 3-$Y$ integral is known, having been evaluated in terms of Wigner 3j symbols \cite{LLQM}.   
Integrals for the present Paper's specific cases of interest are furthermore provided case-by case in e.g. \cite{Mizushima}. 
The first of these integrals occurs in the Stark effect \cite{Stark} ($\pm 1$ selection rule). 
The second in the calculation underlying both Raman spectroscopy \cite{Raman} and Pauling's analysis of the rotation of molecules within crystals \cite{Pauling} ($\pm 2$ selection rule).  
For comparison with the below quadrilateralland working, the first of these integrals has as its nontrivial factor 
$
\int_{-1}^{1}\mP_{\sll^{\prime}}^{\sm^{\prime}}(X)X\mP_{\sll}^{\sm}(X)\d X \mbox{ }  
$
There is then a recurrence relation (\ref{Leg-Rec1})by which $XP_{\sJ}^{\sj}(X)$ can be turned into a linear combination of $P_{\sS^{\prime\prime}}^{\sj^{\prime\prime}}(X)$. 
Finally orthonormality of the associated Legendre functions (\ref{orthog}) can be applied to evaluate it.  
The second of these integrals then requires two uses of the same recurrence relation.  

\mbox{ } 

\noindent RPM Example 1) For 4-stop metroland, the relevant shape integral or perturbed-potential integral is just the l $\rightarrow$ D, m $\rightarrow$ d, 
$\mbox{cos}^2\theta_{\sss\sp} \rightarrow \mbox{cos}^2\theta$ of Atomic Example 2) \cite{AF}.

\noindent RPM Example 2) For triangleland, the relevant shape integral or perturbed-potential integral is just the l $\rightarrow$ S, m $\rightarrow$ j, 
$\mbox{cos}\,\theta_{\sss\sp} \rightarrow \mbox{cos}\,\Theta$ of Atomic Example 2) \cite{+tri}.

\noindent RPM Example 3) Parallels of Atomic Example 1) are given in \cite{AF, +tri} and represent estimations of the RPM Bohr configuration space radius (or, for triangleland, 
Bohr moment of inertia) analogue of Atomic Physics' Bohr radius.  
\noindent One can furthermore view this as part of the Peaking Interpretation of Quantum Cosmology. 
(See e.g. \cite{MR91, EOT, FileR}.) 
Here the lack of universe-measurements rather constrains other means of `interpreting QM'.  

\mbox{ } 

\noindent RPM Example 4) New to the present paper, the relevant quadrilateralland shape integral selection rule for shape quantity $s_8 =$ cos$\,2\chi$ 
as the inserted operator is $\Delta \mI = \Delta \mI_3 = \Delta \mY = 0, \Delta \mn = \pm 1$ or 0.  
It more closely resembles the Stark Effect due to the $\pm 1$ part of its selection rule {\sl despite} how the inserted term itself looks more like 4-stop metroland's.  
(I.e. quadratic rather than linear, like for the Raman Effect.) 
It is a case in which the model being an $N$-a-gon presides over the particle number $N$ being 4, beacuse the Jacobi polynomials themselves are in cos$\,2\chi$.  
Thus this serves as the basic-variable analogue of the Legendre variable, so a `square' insertion is in fact a linear power in the basic `Jacobi variable'.  
It differs by additionally allowing for the non-transition, 0.  
This feature is new to $\mathbb{CP}^2$, arising from the first RHS term in the recurrence relation (\ref{Jac-Rec}) for the Jacobi polynomials. 
This is clearly zero for $\mathbb{S}^{\sp - 1}/\mathbb{R}^{\sp}$/the p-$d$ atom since $\alpha = \beta$ from the Gegenbauer polynomial specialization (\ref{Geg-Rec1}) downward.   
This example therefore unveils a number of good fortunes in the standard atomic version of these calculations, thus serving as a robustness test for the atom.  
\noindent The known surviving terms include the following one that is subsequently used in this paper,
\beq
\langle \Psi_{\sn\,\sI\,\sI_3\,\sY }| \, \mbox{cos}\,2\chi \, |\Psi_{\sn\,\sI\,\sI_3\,\sY }\rangle = 
\frac{Y^2 - \{2\mI + 1\}^2}{\{2\mn + 2\mI + |\mY| + 1\}\{2\mn + 2\mI + |\mY| + 3\}} \mbox{ } .
\label{Quad-Exp}
\eeq
The first two nontrivial transition terms are then  
\beq
\langle \Psi_{\sn - 1\,\sI\,\sI_3\,\sY }| \, \mbox{cos}\,2\chi \, |\Psi_{\sn\,\sI\,\sI_3\,\sY }\rangle = 
\sqrt{\frac{\mn\{\mn + |\mY|\}\{\mn + 2\mI + 1\}\{\mn + 2\mI + |\mY| + 1\}}{\{2\mn + 2\mI + |\mY| + 2\}\{2\mn + 2\mI + |\mY| \}   }  }\frac{1}{2\mn + 2\mI + |\mY| + 1}
\eeq
alongside the  $\mn \rightarrow \mn + 1$ of this.
These results and the subsequent perturbation theory applications can be viewed as further robustness tests for the mathematics and physics of the (arbitrary-dimensional) atom.  

\noindent See e.g. \cite{Stark, Messiah} for atomic counterparts and \cite{AF, +tri, FileR} for 4-stop metroland and triangleland counterparts.

\mbox{ }  

\noindent RPM Example 5) The relevant quadrilateralland scale integrals are
\beq
\langle \rho \rangle = \sqrt{\frac{\hbar}{\omega}}\frac{\Gamma(3\{1 + \sqrt{5}\}/2)}{\Gamma(1 + 3\sqrt{5}/2)} = 2.03\sqrt{\frac{\hbar}{\omega}} \mbox{ } \mbox{ (ground state) } \mbox{ } , 
\eeq
which can be interpreted as a Bohr configuration space radius, and
\beq
\langle \rho^2 \rangle = \{\hbar/\omega\}\{2\mn + 2\Lambda_{n} + 1\} \mbox{ } \mbox{ (for any n) } 
\eeq
by the obvious factorization into scale and shape parts and recurrence relation (\ref{LagRec}).   
%
%
%
See Part III of \cite{FileR} for 4-stop metroland and triangleland counterparts.

\section{Application 1) Expectations and spreads}\label{Exp}

One of us gave these for the `Dragt' shape quantities for triangleland in \cite{+tri} and, in collaboration with Franzen, for 4-stop metroland in \cite{AF}.  
For triangleland, these expectations came out to be zero. 

\noindent Comparison of mean angle (e.g. roughly from the expectation of cos$\,\Theta$) and mode angle (from graphs along the lines of those in \cite{08II}) 
reveals the mean to be larger than the mode, but by not quite as much as occurs radially in the atom.  
This reflects that this case's Gaussianity suppresses the mean-shifting tail more than the radial part of the atom's mere exponential does.  
Expectations and spreads of $\widehat{\Phi}$ are just like for previous Sec as the $\Theta$-integrals trivially cancel in each case.  

\mbox{ } 

\noindent For quadrilateralland, expectation of cos$\,2\chi$ [proportional up to an additive constant to the simplest `Gell-Mann quadratic form' shape quantity (Sec I.16)] 
is given by eq (\ref{Quad-Exp}) and its variance (i.e. measure of spread) by 
$$
\mbox{Var}_{\sn\,\sI\,\sI_3\,\sY }(\mbox{cos$\,2\chi$})= \langle \Psi_{\sn\,\sI\,\sI_3\,\sY }| \, \mbox{cos}^2 2\chi \, |\Psi_{\sn\,\sI\,\sI_3\,\sY }\rangle - 
\langle \Psi_{\sn\,\sI\,\sI_3\,\sY }| \, \mbox{cos}\,2\chi \, |\Psi_{\sn\,\sI\,\sI_3\,\sY }\rangle^2 = 
$$
\beq
4 \times \frac{    \stackrel{\mbox{\normalsize $\mn\{\mn + 2\mI + |\mY| + 1\}\{\mn + 2\mI + 1\}\{\mn + |\mY|\}\{2\mn + 2\mI + |\mY| + 3\}^2\{2\mn + 2\mI + |\mY| + 4\}$ }} 
                   {+ \{\mn + 1\}\{\mn + 2\mI + |\mY| + 2\}\{\mn + 2\mI + 2\}\{\mn + |\mY| + 1\}\{2\mn + 2\mI + |\mY|\}\{2\mn + 2\mI + |\mY| + 1\}^2          }                    }
	  {    \{2\mn + 2\mI + |\mY|\}\{2\mn + 2\mI + |\mY| + 1\}^2\{2\mn + 2\mI + |\mY| + 2\}\{2\mn + 2\mI + |\mY| + 3\}^2\{2\mn + 2\mI + |\mY| + 4\}       }
\eeq
\noindent As limiting cases of particular interest, 1) the ground state has expectation -- 1/3 and variance 2/9. 
\noindent 2) Expectation = $-1/\{2\mn + 3\}\{2\mn + 1\} \longrightarrow 1/4 \mn^2$ as n $\longrightarrow \infty$ and variance $\longrightarrow 1/2$ if I = 0 = Y is kept throughout.  
\noindent 3) On the other hand, if n = I = Y is kept and n is sent to infinity, expectation goes to -- 3/25 and variance to 196/625.

\mbox{ } 

\noindent Finally, Var$\rho = 3.40\hbar/\omega$ for the ground state of the scaled quadrilateral with isotropic HO potential 
-- a particular case of confinedness controlled by the steepness of the well.

\section{Application 3) Emergent time-dependent perturbations in the Semiclassical Approach to Quantum Cosmology}\label{Semicl}  

The specific r-presentation of $N$-a-gonland unapproximated h and l equations are a Hamilton--Jacobi equation (1.17) with quantum correction terms added, 
$$ 
\{\pa_{\sh}S\}^2 - i\hbar \, \pa_{\sh}\mbox{}^2S - 2i\hbar \, \pa_{\sh}S\langle\chi|\pa_{\sh}|\chi\rangle - \hbar^2\big\{  \langle\chi|  \pa_{\sh}\mbox{}^2  |\chi\rangle + 
k(N, d)\mh^{-1}  \langle\chi|\pa_{\sh}|\chi\rangle   \big\} - i\hbar \mh^{-1}k(N, d)\pa_{\sh}S 
$$
\beq
+ \hbar^2 \mh^{-2}\{c(N, d) - \langle\chi|\triangle_{\sll}|\chi\rangle\} + 2 V_{\sh}(\mh) + 2\langle\chi| J(\mh, \ml^{\sfa})|\chi\rangle = 2 E_{\sU\sn\si} \mbox{ } \mbox{ } , 
\eeq
and what is for now a fluctuation equation  
\beq
\{1 - \mP_{\chi}\}    \big\{ - 2i\hbar \, \pa_{\sh}  |\chi\rangle  \pa_{\sh} S - \hbar^2\big\{  \pa_{\sh}\mbox{}^2  |\chi\rangle + k(N, d)\mh^{-1}  \pa_{\sh}  |\chi\rangle  + 
\mh^{-2}\triangle_{l}\}  |\chi\rangle + 2\{V_{\sll}(\ml^{\sfa}) + J(\mh, \ml^{\sfa})\}  |\chi\rangle  \big\} = 0  \mbox{ } .  
\label{l-TDSE-prime}
\eeq
These equations result from those in \cite{Banks, HallHaw} via various specializations in e.g. \cite{Kieferbook, SemiclIII, FileR, ACos2}. 

\noindent The first equation can be cast as a QM-corrected form of the classical energy equation,  
$$ 
\{\Star \mh\}^2 - 2i\hbar \, \Star \mh \langle\chi|\pa_{\sh}|\chi\rangle - \hbar^2\big\{  \langle\chi|  \pa_{\sh}\mbox{}^2  |\chi\rangle + 
k(N, d)\mh^{-1}  \langle\chi|\pa_{\sh}|\chi\rangle   \big\} - i\hbar \mh^{-1}k(N, d)\Star \mh 
$$
\beq
+ \hbar^2 \mh^{-2}\{c(N, d) - \langle\chi|\triangle_{\sll}|\chi\rangle\} + 2 V_{\sh}(\mh) + 2\langle\chi| J(\mh, \ml^{\sfa})|\chi\rangle = 2 E_{\sU\sn\si} \mbox{ } \mbox{ } , 
\eeq
and the second equation into a QM-corrected TDSE, the core of which is

\beq
i\hbar\pa|\chi\rangle/\pa t^{\se\sm(\sW\sK\sB)}  = - \frac{\hbar^2}{\mh^2(t^{\se\sm(\sW\sK\sB)})}\triangle_{\sll}|\chi\rangle 
                                                   + A \mh^2(t^{\se\sm(\sW\sK\sB)}) |\chi\rangle \mbox{ } .  
\label{TDSE2b}
\eeq
[Some omitted correction terms however cause this to depart from being a TDSE.]

\mbox{ } 

\noindent This is a model of the GR Tomonaga--Schwinger equation (given here in relational formulation, i.e. in terms of frame ${F}^{\mu}$ and not shift $\beta^{\mu}$),  
\beq
i\hbar\{\delta/\delta t^{\se\sm} - \big\{\delta{\mF}^{\mu}/\delta t^{\se\sm}\}\widehat{\cal M}_{\mu} \big\}|\chi\rangle = \widehat{H}_{\sll}^{\sG\sR}|\chi\rangle \mbox{ } .  
\label{Tom-Schwi}  
\eeq
(additionally coupled to the quantum momentum constraint equation $\widehat{\cal M}_{\mu}|\chi\rangle = 0$ due to no prior explicit classical reduction being known in this case). 
Such an equation has been considered in more detail in the quantum-cosmological setting by Halliwell and Hawking \cite{HallHaw}.  
The minisuperspace counterpart involves but partial derivatives, no correction term $\dot{F}^{\mu}\widehat{\cal M}_{\mu}$ and no coupled equation.  

\mbox{ } 

\noindent The $|\chi\rangle$ {\sl separates} into a new $t$-or-$\rho$ = h part ${\cal R}$ and the same shape part as in the first half of this paper. 
Thus we identify $|\chi\rangle = ${\cal R}$|\mn\,\mI\,\mI_3\,\mY \rangle$.  
Furthermore N.B. that $\langle \chi| \widehat{O} |\chi \rangle$ involves integration solely over the $l$-space = $\fS(N, d)$.  
Thus we {\sl only} need our pure-shape useful integrals for this application.  

\noindent We now take the classical $t$, solve (\ref{TDSE2b}), and then re-investigate the $\mh$-equation with this approximate knowledge of $|\chi\rangle$.
This is so as to allow the $\ml$-subsystem the opportunity to contribute to a corrected emergent timestandard.  
This is of course along the lines of the ephemeris time procedures outlined in Sec I.22. 
STLRC is based on giving everything the opportunity to contribute but then ditching contributions that turn out to be negligible to the currently requisite accuracy.  
This means quantum emergent time {\sl has} to be different from classical emergent time in principle, since the former has different/additional {\sl quantum} changes contributing to it.  
\cite{QuadI, ARel2, ACos2, FileR} lay this out. 
The part of it we consider in the present paper involves integrating up the h-equation to obtain
\beq
\mbox{\Large $t$}^{\se\sm(\sW\sK\sB)} = \int {2\,\d \mh}\left/\left\{-B \pm \sqrt{B^2 - 4C}\right\}\right. \mbox{ } \mbox{ } \mbox{ for } 
\eeq
\beq
B = - i\hbar\{2 \langle\chi|\pa_{\sh}|\chi\rangle +  \mh^{-1}k(N, d)\} \mbox{ } , \mbox{ } \mbox{ } 
C = -2\{W_{\sh} - \langle\chi| J |\chi\rangle\} + \hbar^2\{   \mh^{-1}k(N, d)\langle\chi|\pa_{\sh}|\chi\rangle - \langle\chi|\pa_{\sh}^2|\chi\rangle + \mh^{-2}c(N, d)\} \mbox{ } .  
\eeq
So what were $\pm$ pairs of solutions to a Hamilton--Jacobi equation (I.17) at the classical level are turned into more distinct complex pairs.
This splitting is mediated by operator-ordering and expectation contributions to first order in $\hbar$. 
One also sees that the second-order contributions are another expectation, another ordering term and one that has one factor's worth of each. 

\noindent \underline{Key 25} This is of the general Machian form [compare the classical counterpart (I.111)]
\beq
\lt^{\se\sm(\sW\sK\sB)} = {\cal F}[\mh, \ml, \d\mh, |\chi(\mh, \ml)\rangle ] \mbox{ } .
\eeq
\noindent Expanding out and keeping up to 1 power of $\hbar$,
\beq
\lt^{\se\sm(\sW\sK\sB)} = \lt^{\se\sm(\sW\sK\sB)}_{(0)} + \frac{1}{2\sqrt{2}}\int\frac{\langle\chi| J |\chi\rangle}{W_{\sh}^{3/2}}\d\mh 
- \frac{i\hbar}{4}  \int   \frac{\d\mh}{W_{\sh}}  \left\{  \frac{k(N, d)}{\mh} + 2\langle\chi|\pa_{\sh}|\chi\rangle \right\} + O(\hbar^2) \mbox{ } .  
\label{QM-expansion}
\eeq
I.e., with comparison with the classical counterpart (I.112) an `expectation of interaction' $\langle J\rangle$ term in place of an interaction term $J$, 
and an operator-ordering term and an expectation term in place of a classical $\ml$-change term.

The simplest case of interaction potential is $J = A\rho^2 \mbox{cos} 2\chi$.  
For this, the ordering term comes out as, in the $A = 0$ case $- i\hbar\{\{n - 1\}/2E_{\sU\sn\si}\}\mbox{ln}\, \mh = - \{i\hbar/E_{\sU\sn\si}\}\mbox{ln}\, \mh$ for quadrilateralland.
For $A > 0$, it comes out as 

\noindent $-i\hbar\{\{n - 1\}/2E_{\sU\sn\si}\}\mbox{ln}(\mh/\sqrt{E - A\mh^2}) = -\{i\hbar/E_{\sU\sn\si}\}\mbox{ln}(\mh/\sqrt{E - A\mh^2})$ for quadrilateralland. 
N.B. this term does not involve any kind of coupling to the l-equation, unlike the next three terms considered.  

\noindent Via (\ref{Quad-Exp}), the $J$ expectation term comes out as  
\beq
\frac{1}{2\sqrt{2}}
\left\{  
\frac{Y^2 - \{2I + 1\}^2}{\{2\mn + 2\mI + |\mY| + 1\}\}\{2\mn + 2\mI + |\mY| + 3\}}
\right\}
\frac{1}{\sqrt{A}}
\left\{
\frac{B}{A}
\right\}
\left\{
\frac{\sqrt{A} \mh}{\sqrt{E_{\sU\sn\si} - A \mh^2}} - \mbox{arcsin}\{\sqrt{A/E_{\sU\sn\si}} \mh\} 
\right\}
\eeq
$B/A = \{K_2 - K_1\}/\{K_1 + K_2\}$, so this factor is an approximate contents homogeneity smallness.


\noindent The other first-order expectation term, $\langle \pa_{\sh} \rangle$, is
$$
-\frac{i\hbar}{2}\int\frac{\d \mh}{W_{\sh}}{\cal R}^*\frac{\d {\cal R}}{\d t^{\se\sm(\sW\sK\sB)}}\frac{\d t^{\se\sm(\sW\sK\sB)}}{\d \mh} = 
-\frac{1}{2\sqrt{2}}\int\frac{\d h}{h^2W_{\sh}^{3/2}}{\cal R}^*{\cal R}\{2\hbar^2k\{k + n - 1\} + A\mh^4\} \hspace{3.15in}
$$
\beq
\hspace{1.6in} = \frac{- \{4\hbar^2k\{k + n - 1\}A + 3 E_{\sU\sn\si}^2\}\mh + E_{\sU\sn\si}\mh^3}{4\sqrt{2}E_{\sU\sn\si}\sqrt{E_{\sU\sn\si} - A\mh^2}A}
+ \frac{3E_{\sU\sn\si}}{4\sqrt{2}A^{3/2}}\mbox{arctan}\left(\frac{\sqrt{A}\mh}{\sqrt{E_{\sU\sn\si} - A\mh^2}}\right) \mbox{ } .  
\eeq

\noindent The higher-order derivative counterpart of the preceding, which occurs to second order is also analytically computible, coming out as proportional to

\noindent
$$
\frac{1}{6\sqrt{2}}\frac{-3AE_{\sU\sn\si}^3\mh^6 + \{9E_{\sU\sn\si}^4 + 4A\Lambda_n\{4A\Lambda_n + 3E_{\sU\sn\si}^2\}\}\mh^4 - 8AE_{\sU\sn\si}\Lambda_n^2\mh^2 - 2\Lambda_n^2E_{\sU\sn\si}^2}
{E_{\sU\sn\si}^3\mh^3\sqrt{E_{\sU\sn\si} - A\mh^2}} - \frac{3E_{\sU\sn\si}}{2\sqrt{2}}\mbox{arctan}\left(\frac{\sqrt{A}\mh}{\sqrt{E_{\sU\sn\si} - A\mh^2}}\right)
$$
\beq
- {i\hbar}
\left\{
\frac{2\Lambda_n}{\mh^2} + \frac{A\Lambda_n}{E_{\sU\sn\si}\{E_{\sU\sn\si} - A\mh^2\}} + \frac{E_{\sU\sn\si}}{E_{\sU\sn\si} - A\mh^2} - 
\frac{2A\Lambda_n}{E_{\sU\sn\si}^2}\mbox{ln}\,\mh  + \left\{1 -  \frac{A\Lambda_n}{E_{\sU\sn\si}^2}  
\right\}
\mbox{ln}\{A\mh^2 - E_{\sU\sn\si}\} 
\right\} + \mbox{const} \mbox{ } .
\eeq
\noindent Note that the shape--scale TISE possesses {\sl exact} solutions of what the semiclassical approach's TDSE merely approximates. 
So Secs \ref{sep}--\ref{scale} will eventually furnish {\sl tests} for whether the main, and then smaller, regime choices done in the Semiclassical Approach are consistent.  

\noindent Also note that there is a widespread prejudice in Semiclassical Quantum Cosmology that expectation terms are always small.  
However, counterexamples to this were given in \cite{SemiclIII} for triangleland (expectation of l-Laplacian term when the wavefunction is an eigenvector of the Laplacian).
In the present Paper we have shown furthermore that in some regions of configuration space the expectations of the interaction $J$, of $\pa_{\sh}$ and of $\pa^2_{\sh}$ also blow up. 
Thus the negligibility of expectation terms suffers from a Global Problem.  
Moreover, having to keep average terms spells the ends to any claims of Semiclassical Quantum Cosmology being an analytically-tractable subject. 
It would then not only be a numerical subject due to its many-term equations but also be of integro-differential form reminiscent of the Hartree--Fock \cite{AtF}                
approximate formulation of Atomic and Molecular Physics.  
In this case, around 1930 or so, the inclusion of expectation terms was found to be {\sl highly necessary} 
in order to at all accurately reproduce atomic and molecular spectra observations from one's Quantum Theory.
See \cite{FileR, SemiclIV} for more. 

\mbox{ } 

\noindent In terms of emergent {\it `rectified time'} $t^{\se\sm(\sr\se\sc)}$ given by $\pa/\pa t^{\se\sm(\sr\se\sc)} := \mh^2\pa/\pa t^{\se\sm(\sW\sK\sB)}$ or, 
in shorthand, $\mbox{\textcircled{$\star$}}$ := $\mh^2\Star$, the l-equation is now cleaner, its $t$-dependence now being in line with basic Physics' TDSE: 
\beq
i\hbar\pa|\chi\rangle/\pa t^{\se\sm(\sr\se\sc)}  = - \{\hbar^2/2\} \triangle_{\sll} |\chi\rangle + A\rho^4(t^{\se\sm(\sr\se\sc)}) |\chi\rangle \mbox{ } .  
\label{rec-TDSE}
\eeq
It then makes sense \cite{ACos2} to recast the h-equation in this same time (recollect the classical-level motivation for $t^{\se\sm(\sJ\sB\sB)}$: the simplifier of equations of motion.  
It just now turns out that QM implies a conformally-related rectified time is to be used at the quantum level instead.  
The h-equation is then
$$
\{\Rec \, \mbox{ln} \, h\}^2 - 2i\hbar\langle \chi |  \Club  | \chi \rangle - \hbar^2\{ \langle \chi | \Spade^2 | \chi \rangle + k(N, d) \langle \chi | \Spade | \chi \rangle \} 
$$
\beq
- i\hbar k(N, d) \Rec \, \mbox{ln} \, \mh + \hbar^2\{ k(\xi) - \langle \triangle_{\sll} \rangle = 
2\{E^{\sr\se\sc} -  V_{\mh}^{\sr\se\sc} - \langle \chi | V_{\sll}^{\sr\se\sc} | \chi \rangle - \langle \chi | J^{\sr\se\sc} | \chi \rangle \}\mbox{ } . 
\eeq
for  $\Club    := \Rec  - \Rec  \ml\pa_{\sll}$  and $\Spade := \Club/\Rec  \, \mbox{ln} \, \mh(t^{\se\sm(\sr\se\sc)})$.  
Integrating this gives 
\beq
\mbox{\Large $t$}^{\se\sm(\sr\se\sc)} = \int {2\,\d \mh}\big/{\mh^2\big\{  -B \pm \sqrt{B^2 - 4C}\big\}}    \mbox{ } .  
\eeq
\noindent Expanding out and keeping up to 1 power of $\hbar$,
\beq
\lt^{\se\sm(\sr\se\sc)} = \lt^{\se\sm(\sr\se\sc)}_{(0)} + \frac{1}{2\sqrt{2}}\int\frac{\langle\chi| J |\chi\rangle}{\mh^2W_{\sh}^{3/2}}\d\mh 
-\frac{i\hbar}{4}  \int   \frac{\d\mh}{\mh^2 W_{\sh}}  \left\{  \frac{k(N, d)}{\mh} + 2\langle\chi|\pa_{\sh}|\chi\rangle \right\} + O(\hbar^2) \mbox{ } .
\label{QM-expansion-rec}
\eeq
\noindent The ordering term to the rectified time then comes out as, in the  $A = 0$ case, $i\hbar\{n - 1\}/2E_{\sU\sn\si}\mh^2 = i\hbar/E_{\sU\sn\si}\mh^2 $ for quadrilateralland.
For $A > 0$, it is $i\hbar\{\{n - 1\}/2E_{\sU\sn\si}\}\{1/2\mh^2 - \{A/E_{\sU\sn\si}\}\mbox{ln}(\mh/\sqrt{E - A\mh^2})\} =
\{i\hbar/E_{\sU\sn\si}\}\{1/2\mh^2 - \{A/E_{\sU\sn\si}\}\mbox{ln}(\mh/\sqrt{E - A\mh^2})\}$ for quadrilateralland.
%
%
\noindent The $\langle J \rangle$ correction term to the rectified time is, via (\ref{Quad-Exp}) again, 
\beq
\frac{1}{2\sqrt{2}}
\left\{  
\frac{Y^2 - \{2I + 1\}^2}{\{2\mn + 2\mI + |\mY| + 1\}\}\{2\mn + 2\mI + |\mY| + 3\}}
\right\}
\left\{
\frac{B}{E_{\sU\sn\si}}
\right\}
\frac{\mh}{\sqrt{E_{\sU\sn\si} - A \mh^2}} \mbox{ } .  
\eeq
\noindent
The other first-order expectation term, $\langle \pa_{\sh} \rangle$, is now 
\beq
- 
\frac{E_{\sU\sn\si}^2\mh^2 + 2\hbar^2k\{k + n - 1\}\{2A\mh^2 - E\sU\sn\si\}}{2\sqrt{2}\,E_{\sU\sn\si}^2\mh\sqrt{E - A\mh^2}} 
+ \frac{1}{2\sqrt{2A}}
\mbox{arctan}
\left(
\frac{\sqrt{A}\mh}{\sqrt{E_{\sU\sn\si} - A\mh^2}}
\right) \mbox{ } .
\eeq


\noindent Its higher-order derivative counterpart that occurs to second order is now proportional to

\noindent
$$
\mbox{const} + \frac{1}{5\sqrt{2}}\frac{A\{5E_{\sU\sn\si}^4 + 16A\Lambda_n^2 + 20AE_{\sU\sn\si}^2\Lambda_n\}\mh^6 - 2AE_{\sU\sn\si}\Lambda_n\{5E_{\sU\sn\si}^2 + 4A\Lambda_n\}\mh^4 - 2AE_{\sU\sn\si}^2\Lambda_n^2\mh^2 - \Lambda_n^2E_{\sU\sn\si}^3}{E_{\sU\sn\si}^4\mh^5\sqrt{E_{\sU\sn\si} - A\mh^2}} 
$$

\noindent
\beq
-\frac{\sqrt{A}}{\sqrt{2}}\mbox{arctan}\left(\frac{\sqrt{A}\mh}{\sqrt{E_{\sU\sn\si}\!\!-\!\! A\mh^2}}\right)
-{i\hbar}
\left\{
\frac{\Lambda_n}{E_{\sU\sn\si}\mh^4}\!\!+\!\!\frac{A\Lambda_n}{E_{\sU\sn\si}^2\mh^2}\!\!+\!\!\left\{\frac{A\Lambda_n}{E_{\sU\sn\si}^2}\!\!-\!\!1\right\}
\frac{A}{E_{\sU\sn\si}\!\!-\!\!A\mh^2}\!\!+\!\!\frac{4A}{E_{\sU\sn\si}}\mbox{ln}\,\mh\!\!-\!\!\frac{2A}{E_{\sU\sn\si}}
\mbox{ln}\left( - E_{\sU\sn\si}\!+\!A\mh^2\right)\right\} .
\eeq

\noindent
More advanced cases of coupled h- and l-equation schemes are considered in \cite{SemiclIII, FileR, ACos2, SemiclIII}, 
albeit just for 3-stop metroland and a few triangleland workings so far.

\mbox{ } 

\noindent N.B. the $\langle J \rangle$, $\langle \pa_{\sh} \rangle$, $\langle \pa^2_{\sh} \rangle$ terms are all contributions to/mechanisms for backreaction. 
Morevover, the $\langle J \rangle$ integral backreaction mechanism allowed for quadrilateralland is forbidden by symmetry/selection rules in the case of the triangle.  
This triangle to quadrilateral difference directly reflects the 0 selection rule that is afforded by the Jacobi polynomials but not by their (Gegenbauer and) Legendre specializations.

\section{\NSI}\label{NSI}

Armed with the present Paper's wavefunctions, let us now complete the \NSI \cite{HP86, UW89} evaluation of 
\beq
\mbox{Prob}(\mbox{Region R}) \propto \int_{\sR}|\Psi_{\sn\,\sI\,\sI_3\,\sY}|^2 \d\Omega \mbox{ } . 
\eeq
as set up in Paper I and now labelled with this paper's quartet of quantum numbers.  

\mbox{ }  

\noindent Example 1)
\beq
\mbox{Prob}(\epsilon\mbox{-collinear}) \propto \int_{\sC_{\epsilon}}|\Psi_{\sn\,\sI\,\sI_3\,\sY}|^2\d\Omega
\eeq
for the $\mC_{\epsilon}$ supplied in (I.147). 
Thus, taking the Gibbons--Pope type coordinates of this series of papers,
$$
\mbox{Prob}(\epsilon\mbox{-collinear}) \propto 
\int_{\psi = 0, \pi - \epsilon, 2\pi - \epsilon, 3\pi - \epsilon, 4\pi - \epsilon}^{\epsilon, \pi - \epsilon, 2\pi - \epsilon, 3\pi - \epsilon, 4\pi}
\int_{\phi = 0, \pi - \epsilon, 2\pi - \epsilon}^{\epsilon, \pi - \epsilon, 2\pi}
\int_{\beta = 0}^{\pi}
\int_{\chi = 0}^{\pi/2}
|\Psi_{\sn\,\sI\,\sI_3\,\sY}(\chi, \beta, \phi, \psi)|^2
\mbox{sin}^3\chi\,\mbox{cos}\,\chi\,\d\chi\, \mbox{sin}\,\beta\,\d\beta\, \d\phi \, \d\psi
$$
\beq
\propto \epsilon^2 \mbox{ } \mbox{ i.e. the cross-section of the `2-lune'} \mbox{ for all at-least $U(1) \times U(1)$ symmetric wavefunctions, i.e. Y = 0 = I$_3$ } .
\eeq
This concerns highly non-uniform states according to the demo(4) measure of uniformity (c.f. Sec I.17).
On the other hand, one of the first Y = 0, I$_3 \neq$ 0 solutions gives the additional factor $\{1 - 2I_3^2\epsilon\}$ i.e. a small decrease.  
This shows that adding `isospin' has the same-sign effect as adding relative dilational momentum to 4-stop metroland or mixed shape momentum to triangleland \cite{FileR}.  

\noindent Example 2) Using the $\mT_{\epsilon}$ supplied in Appendix I.A,
$$
\mbox{\scriptsize Prob($\epsilon$-close to a +43 triangle)} \propto \int_{\sT_{\epsilon}}|\Psi_{\sn\,\sI\,\sI_3\,\sY}|^2\d\Omega
= 
\int_{\psi = 0}^{4\pi}
\int_{\phi = 0}^{2\pi}
\int_{\beta = \pi - \epsilon}^{\pi}
\int_{\chi = 0}^{\pi/2}
|\Psi_{\sn\,\sI\,\sI_3\,\sY}(\chi, \beta, \phi, \psi)|^2
\mbox{sin}^3\chi\,\mbox{cos}\,\chi\,\d\chi\, \mbox{sin}\,\beta\,\d\beta\, \d\phi \, \d\psi
$$
\beq
\propto \epsilon \mbox{ } \mbox{ i.e. the width of the belt} \mbox{ for all at-least $SU(2) \times U(1)$ symmetric wavefunctions, i.e. Y = 0 = I = I$_3$} .
\eeq
\noindent Example 3) Using the $\mS_{\epsilon}$ supplied in Appendix I.A, 
$$
\mbox{Prob}(\epsilon\mbox{-close to the 1243-labelled square}) \propto \int_{\sS_{\epsilon}}|\Psi_{\sn\,\sI\,\sI_3\,\sY}|^2\d\Omega
= 
$$
\beq
\int_{\psi = \pi - \epsilon}^{\pi + \epsilon}
\int_{\phi = 0, 2\pi - \epsilon}^{\epsilon, 2\pi}
\int_{\beta = \pi/2 - \epsilon}^{\pi/2 + \epsilon}
\int_{\chi = \pi/4 - \epsilon}^{\pi/4 + \epsilon}
|\Psi_{\sn\,\sI\,\sI_3\,\sY}(\chi, \beta, \phi, \psi)|^2
\mbox{sin}^3\chi\,\mbox{cos}\,\chi\,\d\chi\, \mbox{sin}\,\beta\,\d\beta\, \d\phi \, \d\psi
\eeq
\beq
\propto \epsilon^4 \mbox{ } \mbox{ i.e. the size of the `4-box'} \mbox{ for the ground state n = 0 = I = I$_3$ = Y} .
\eeq
This one is a question of maximal uniformity, in fact more sharply defined than the demo(4) measure can provide, since the squares are not the only configurations that maximize that.  
Note that these are not substantially more probable than for other regions of configuration space of the same size. 
Thus there is {\sl not} a big peak on high uniformity, unlike in one form of Barbour's conjecture \cite{EOT}. 
This concurs with Sec 6's analysis.  

\mbox{ }  

\noindent For comparison, GR Cosmology \NSI calculations can be found e.g. in \cite{HP86}.

\section{Conclusion} \label{Concl2}

\subsection{Quantization via use of geometrical methods}

\noindent We provided kinematical quantization of the relational quadrilateral. 
The pure-shape version is based on the quadrilateral shape space's $\mathbb{CP}^2$'s isometry group $SU(3)/\mathbb{Z}_3$ 
and the linear space of Gell-Mann quadratic forms provided by Paper I.  
This is a coherent extension of triangleland's kinematical quantization, but only once one has taken into account the $\mathbb{R}^3$ vector to Pauli matrix map 
by which triangleland's shape quantities form Sec I.16's space IHP($\mathbb{C}^2$, 2).  
\noindent Using $\mathbb{C}^{n}$ for the linear space is geometrically natural too, but leads to absolutist rather than relational physics.
For triangleland, 

\noindent IHP($\mathbb{C}^2$, 2) = $\mathbb{R}^3$ has more minimal dimension than $\mathbb{C}^2$, whilst for quadrilateralland, this minimality is reversed.
(Isham \cite{I84} suggested, but did not oblige, dimensional minimality for the linear space to be involved in kinematical quantization.)
\noindent The scaled version of kinematical quantization is based on the extension of this to the cone over $\mathbb{CP}^2$.

We then provided time-independent Schr\"{o}dinger equations for pure-shape and scaled quadrilateralland.  
The first of these can build upon, after quantum-cosmological conformal-term adjustment, MacFarlane's work \cite{MacFarlane}.  
In the free case, it separates in Gibbons--Pope type coordinates. 
The second of these shape--scale splits into a different adjustment of the preceding pure-shape problem and a scale equation 
that just maps to the Bessel and associated Laguerre equations for the free and isotropic HO problems respectively.

\subsection{Interplay with Atomic/Molecular Physics, Particle Physics and Shape Geometry}

\noindent Next, we considered useful integrals -- 
the analogue of those used in the atomic Stark Effect and for expectations and spreads of radial and shape operators in Atomic and Molecular Physics.
In the quadrilateralland context, these can be used for time-independent perturbations about exact solutions, and time-dependent perturbations as useful in the Semiclassical Approach 
to the PoT and Quantum Cosmology.  
Furthermore, expectations and spreads of scale and shape operators are now a concrete means of carrying out the Peaking Interpretation of Quantum Cosmology \cite{MR91, EOT, FileR}.  

\mbox{ } 

\noindent The QM of $\mathbb{CP}^2$ serves as a robustness test of the ($k$-dimensional) rotor and atom problems as follows.

\noindent A) this problem's orbitals can no longer be labelled by an obvious surrounding Cartesian space's axes (i.e. $p_x, p_y, p_z, ...$ for the atom).  
Instead, it has an octet of `$p$-orbitals' that bear the same group-theoretic relations as in Gell-Mann's eightfold way in Particle Physics.
Two things have happened here.  

\noindent 1) The Cartesian space axes once again generalize in this way via equivalence to the Pauli matrices [$SU$(2) adjoint rep].

\noindent 2) $\mathbb{CP}^2$ has in excess of the degeneracy of orbitals possessed by that dimension's maximally symmetric shape space, $\mathbb{S}^4$ 
(e.g. by 8 to 5 for the $p$-orbitals).

\noindent B) The $\pm 1$ selection rule problem now comes with a non-transition term (0 selection rule) absent from the $k$-dimensional atom. 
This is a direct consequence of the Jacobi polynomials being more general than the Gegenbauer polynomials [compare (\ref{Jac-Rec}) and (\ref{Geg-Rec1})].  
By it, the result concerning the second-order nature of the first perturbation terms for the rotor Stark effect does not carry over to $\mathbb{CP}^2$.    

\mbox{ } 

\noindent We used the complex-projective chopping board of Paper I as a back-cloth for discussing the wavefunctions. 
This is in parallel to the use of spherical blackboards in \cite{+tri, AF, FileR}, the triangleland case of which origninates in Kendall's work on Shape Statistics.
What RPM's give back to that subject are questions concerning Geometrical Statistics and analogues of Shape Statistics itself for GR, QM and Quantum Gravity.

Cones over complex projective spaces (and quotients of complex projective spaces and cones over those two) 
have featured in the String Theory literature e.g. as models of orbifolds \cite{WCP}.  
Connections between these and Mechanics have long been pointed out by Atiyah.

\subsection{$N$-a-gonland generalization of this Paper} 

\noindent As the present Paper makes clear, quadrilateralland is far closer to the general $N$-a-gon in terms of resultant mathematics, so the present paper is also the true gate 
to the general $N$-a-gon. 
This SSec's considerations also lead to e.g. i) more general robustness studies along the lines of \cite{KR89}. 
ii) Large-$N$ considerations such as Statistical Mechanics.
iii) A Shape Statistics approach \cite{Kendall8489, Kendall} to Records Theory \cite{AStats}. 
iv) Study of the behaviour in the large-$N$ limit. 

\noindent Now the kinematical quantization involves Isom($\mathbb{CP}^{k}$) = $SU(k + 1)/\mathbb{Z}_{k + 1}$ and IHP($\mathbb{C}^n$, 2).  
As regards nonminimality, $N$-a-gonland takes after quadrilateralland, since dim(Adj($SU$(n)) = $n^2 - 1 > 2n$ = dim($\mathbb{C}^n$) for $n > 2$ (i.e. $N > 3$).

The argument for conformal operator ordering is certainly general enough to hold for all $N$-a-gonlands.  
The resultant TISE is given in \cite{FileR} in complex Fubini--Study coordinates. 
Some further work on this equation is given by MacFarlane in \cite{MF03b}. 
In particular, there continues to be an analogue of the radial $\chi$ part that separates out, 
and this continues to map to the hypergeometric equation and thus to give the Jacobi polynomials. 
The other separated-out part at this stage is a generalized Euler angle part consisting of the $SU(N - 2)$ analogue of the $SU(2)$ Wigner D-function. 
Here for now we do not know the extent to which the requisite analogy has been tabulated yet. 
This reason and making the new points without overly complicating the calculations and the presentation is why for now we stop at quadrilateralland rather than solving for any $N$-a-gon.  
MacFarlane also explicitly treats the whole TISE for $\mathbb{CP}^3$ (i.e. for us, pentagonland) using the counterpart of the Gibbons--Pope type coordinates for this case.

Also, via (\ref{A1}, \ref{A2}) and Sec 2, 
\beq
\mbox{(Energy of $k$th eigenstate of $N$-a-gonland) , } \mbox{ } \fE = 2\hbar^2\{k\{k + n - 2\} + n\{n - 1\}\{n - 2\}/\{2n - 3\}\}  
\mbox{ } , k \, \in \, \mathbb{N}_0 \mbox{ } , 
\eeq
with degeneracies
\beq
D(k, N - 2)  = \{N - 2\}\{N - 2 + 2k\}
\left\{
\frac{(N + k - 3)!}{(N - 2)! \, k!}
\right\}^2 \mbox{ } .
\label{clippe} 
\eeq
In particular, one has
1  ground state                                        `$s$-orbital', 
\{$N$ -- 2\}$N$ = $n^2 - 1$  first excited state        `$p$-orbitals' and 
\{$N$ -- 1\}$^2$\{$N^2$ -- 4\}/4 second excited state      `$d$-orbitals'. 
Thus, interpreting MacFarlane's \cite{MF03b} in pseudo-atomic whole-universe terms, the 6-$d$ $\mathbb{CP}^{3}$ pentagonland QM has 15 `$p$-orbitals' and 84 `$d$-orbitals', the 8-$d$ 
$\mathbb{CP}^{4}$ hexagonland QM has 24 `$p$-orbitals' and 200 `$d$-orbitals', and that the 8-$d$ $\mathbb{CP}^{5}$ heptagonland QM has 35 `$p$-orbitals' and 405 `$d$-orbitals'.

As regards comparing HO potential terms and $\mathbb{CP}^k$ harmonics, $\mathbb{CP}^k$'s first harmonic has dimension $\{k + 1\}^2 - 1$ i.e. the adjoint representation's i.e. 
that of the $SU(k + 1)$ itself.   
The HO's are pure-symmetric, so there are $\{k + 2\}\{k + 1\}/2$ -- 1 anisotropic modes among these, so they only cover part of the possible first harmonics.
Contrast with the $N$-stop metrolands for which these are {\sl all} of the second harmonics \cite{AF}.  
The allness is due to lack of spatial antisymmetry in metroland -- it is about what is omitted principally, which is the $\{k + 1\}k/2$ antisymmetric polynomials.  
This is because jth order harmonics are a basis for jth order polynomials but these can include antisymmetric polynomials for spatial dimension $>$ 
1 and these are not among the HO potentials.

Our useful $s_8$ = cos$\,2\chi$ insertion $\chi$-integral has a clear $N$-a-gonland counterpart.
The $\chi$ part of this is no harder than in the present paper, though the remainder is less well-known for the analogues of the Wigner D-functions.

The scaled $N$-a-gonland's scaled part continues to separate out and give an equation of the same general form as for quadrilateralland 
(or, for that matter, any other RPM): eq (\ref{Nagon-chi}).

\subsection{Problem of Time Applications in this Paper} 

\noindent 1) We considered quantum \K beables for the quadrilateral by aligning them with the kinematical quantization algebra, so that they are the 8 $SU(3)$ 
generators and the 8 Gell-Mann quadratic forms. 
The most natural language for expressing all 16 of these at once is in terms of Gibbons--Pope type intrinsic coordinates.  

\noindent 2) We considered the Machian version of Semiclassical Approach around the quantum Frozen Formalism Problem. 
In particular, we consider Machian correction terms to the zeroth approximation (itself not Machian) for the WKB time.  
Namely, we considered an operator-ordering term that can be treated decoupled from the quantum l-physics of shapes and three types of backreaction terms that 
do require solving the quantum l-TDSE.

\noindent 3) This paper's wavefunctions complement Paper I's characterization of physical propositions in terms of geometrically-simple regions of configuration space 
so as to be able to conclude \NSI calculations.
This included consideration of quantum-cosmologically relvant questions concerning maximal and minimal uniformity.  

\noindent Paper III will contain each of Histories Theory and Records Theory for classical and quantum quadrilateralland.  
%
%
Paper IV will combine these with the present Paper's Machian Semiclassical Approach as a more advanced example of the program in \cite{H03, H09, AHall, FileR, CapeTown12}.  
This is a useful advance for the reasons given in the last paragraph of Paper I's Conclusion. 
It includes promoting the semiclassical quantum \K beables resolution to a quantum Dirac beables one.

\mbox{ }

\noindent {\bf Acknowledgements}: E.A.: I thank those close to me for being supportive of me whilst this work was done. 
Professors Don Page and Gary Gibbons for teaching me about $\mathbb{CP}^2$.
Dr Julian Barbour for introducing me to RPM´s.  
Mr Eduardo Serna for discussions.  
Professors Marc Lachi\`{e}ze-Rey, Malcolm MacCallum, Don Page, Reza Tavakol and Jeremy Butterfield for support with my career.  
E.A.'s work was funded by a grant from the Foundational Questions Institute (FQXi) Fund, 
a donor-advised fund of the Silicon Valley Community Foundation on the basis of proposal FQXi-RFP3-1101 to the FQXi, whilst employed at APC Universit\'{e} Paris Diderot in 2012.   
Thanks also to Theiss Research and the CNRS for administering this grant.

\appendix\section{The $\mathbb{CP}^{\sN}$ eigenspectrum}\label{CPN}

\noindent From Berger et al \cite{BGM71},\footnote{This includes correcting a typo in the latter and making the following minor clarification. 
For $k = 0$, in the conceptual form, the reasoning is that $\mbox{\large$($}\stackrel{\mbox{\tiny N -- 1}}{\mbox{\tiny --1}}\mbox{\large$)$}$ 
is not a possible choosing process and therefore zero. 
Thus $D(0,\sN) = \mbox{\large$($}\stackrel{\mbox{\tiny N -- 1}}{\mbox{\tiny 0}}\mbox{\large$)$}^2 - 0 = 1$ as indeed befits ground states.}
consideration of the eigenvalue problem for $\mathbb{CP}^{\sN}$ (i.e. $\{\triangle + {\cal E}\}u = 0$ so as to compare with our own convention for the free TISE),
\noindent 
\beq
\mbox{($k$th eigenvalue of $\mathbb{CP}^{\sN}$) , } \mbox{ } {\cal E}(k, \mN) = 4k\{\mN + k\} \mbox{ } , k \, \in \, \mathbb{N}_0 \mbox{ } , \mbox{ } \mbox{ and } 
\label{A1}
\eeq
\beq
\mbox{(Degeneracy of $k$th eigenstate of $\mathbb{CP}^{\sN}$) , } \mbox{ } {\cal D}(k, \mN) 
\stackrel{\mbox{\scriptsize conceptually}}{=} \mbox{\LARGE$($}\stackrel{\mbox{\scriptsize N + $k$}}{\mbox{\scriptsize $k$}}\mbox{\LARGE$)$}^2 
                                                    -  \mbox{\LARGE$($}\stackrel{\mbox{\scriptsize N + $k$ -- 1}}{\mbox{\scriptsize $k$ -- 1}}\mbox{\LARGE$)$}^2 \mbox{ } 
\eeq
(from relating the eigenspaces to spaces of homogeneous polynomials whose dimension is elementarily computible), 
and which then simplifies to the more computationally useful form
\beq
{\cal D}(k, \mN) = \mN\{\mN + 2k\}
\left\{
{(\mN + k - 1)!}/{\mN! \, k!}
\right\}^2 \mbox{ } .
\label{A2} 
\eeq
We note that eq. 65 ii) of Macfarlane \cite{MF03b} has typos in it and should be replaced by 
\beq
\mbox{dim(2, 0, 2)} = {\cal D}(2, \mN) = \mN\{\mN + 1\}^2\{\mN + 4\}/4 \mbox{ } 
\eeq
(to convert between notations, our N is his $n$).

\section{Jacobi polynomials}\label{JP}

\noindent The Jacobi polynomials \cite{AS, CHMFSzego, GrRy, Ismail} $\mP^{(\alpha, \beta)}_{\sn}(x)$ are the terminating series solutions that solve 
\beq
\{1 - x^2\}y^{\prime\prime} + \{\beta - \alpha - \{\alpha + \beta + 2\}x\}y^{\prime} + \mn\{\mn + \alpha + \beta + 1\}y = 0 \mbox{ } , \mbox{ } \mbox{ }
\alpha, \mbox{ } \beta > -1 \mbox{ } ,
\label{de}
\eeq
which is the {\it hypergeometric equation} (under the map $x = 1 - 2\eta^2$), this being the most general second-order linear o.d.e. 
in the complex plane to possess three simple poles.
It includes the Gegenbauer alias ultraspherical polynomials as a special subcase ($\alpha = \beta$), with both the Legendre polynomials 
($\alpha = 0 = \beta$) and the Tchebychev polynomials of the first kind ($\alpha = -1/2 = \beta$) as special subcases of that \cite{AS}. 

\noindent The Jacobi polynomials are standardized according to $\mP^{(\alpha, \beta)}_{\sn}(1) = \big( \stackrel{\sn + \alpha}{\mbox{\scriptsize n}}\big)$.  

\noindent By recasting (\ref{de}) in the Sturm--Liouville form 
\beq
\big\{\{1 - x\}^{\alpha + 1}\{1 + x\}^{\beta + 1}y^{\prime}\big\}^{\prime} + n\{n + \alpha + \beta + 1\}\{1 - x\}^{\alpha}\{1 + x\}^{\beta}y = 0  
\mbox{ } ,
\eeq
one can read off that the weight function is $\{1 - x\}^{\alpha}\{1 + x\}^{\beta}$.  
The orthonormality relation is then 
\beq
\int_{-1}^{+1}\{1 - x\}^{\alpha}\{1 + x\}^{\beta}\mP^{(\alpha, \beta)}_{\sm}(x)\mP^{(\alpha, \beta)}_{\sn}(x)\d x = 
\frac{2^{\alpha + \beta + 1}}{2\mn + \alpha + \beta + 1} 
\frac{ \Gamma(\mn + \alpha + 1)\Gamma(\mn + \beta + 1)}{\Gamma(\mn + \alpha + \beta + 1) \mn! }\delta_{\sm\sn} \mbox{ } \mbox{ for } \mbox{ } \mR\me\,\alpha, \mR\me\,\beta > -1 \mbox{ } .
\eeq 

\noindent We also need the following recurrence relation: 
$$
2\{\mn + 1\}\{\mn + \alpha + \beta + 1\}\{2\mn + \alpha + \beta\}\mP^{(\alpha, \beta)}_{\sn + 1}(x) = 
\{2\mn + \alpha + \beta + 1\}
\{    \{\alpha^2 - \beta^2\} + \{2\mn + \alpha + \beta\}\{2\mn + \alpha + \beta + 2\} x    \}\mP^{(\alpha, \beta)}_{\sn}(x) 
$$
\beq
\hspace{3in} - 2\{\mn + \alpha\}\{\mn + \beta\}\{2\mn + \alpha + \beta + 2\}\mP^{(\alpha, \beta)}_{\sn - 1}(x)\mbox{ } . 
\label{Jac-Rec}
\eeq

\mbox{ } 

\noindent{\bf For useful comparison: spherical harmonics and the (associated) Legendre equation}

\mbox{ } 

\noindent The associated Legendre equation is 
\beq
\{1 - X^2\}Y_{,XX} - 2XY_{,X} + \{\mJ\{\mJ + 1\} - \mj^2\{1 - X^2\}^{-1}\}Y = 0 \mbox{ } .
\label{Leg}
\eeq
This comes from the $\theta$ part of the spherical harmonics p.d.e. (X = cos$\,\theta$; the $\phi$ part gives just SHM).

It is solved by the associated Legendre functions $P_{\sJ}^{|\sj|}(X)$ for J $\in \mathbb{N}_0$, j $\in \mathbb{Z}$, $|\mj| \leq$ J.
We use the standard convention that
\beq 
P^{\sj}_{\sJ}(X) = \{-1\}^{\sj}\{1 - X^2\}^{\frac{\sj}{2}}\frac{\d^{\sj}}{\d X^{\sj}}
\left\{
\frac{1}{2^{\sJ}\mJ !}\frac{\d^{\sJ}}{\d X^{\sJ}}\{X^2 - 1\}^{\sJ}
\right\} \mbox{ } , 
\eeq
by which 

\noindent
\beq
\left\{ 
\sqrt{    \frac{2\mJ + 1}{2}  \frac{    \{\mJ - |\mj|\}!    }{    \{\mJ + |\mj|\}!    }    }
P^{|\sj|}_{\sJ}(X)
\right\}
\label{orthog}
\eeq
is a complete set of orthonormal functions for $X \in$ [--1, 1].  
We also require the recurrence relation \cite{GrRy, AS} 
\beq
XP^{|\sj|}_{\sJ}(X) = 
\frac{\{\mJ - |\mj| + 1\}P^{|\sj|}_{\sJ + 1}(X) + \{\mJ + |\mj|\}P^{|\sj|}_{\sJ - 1}(X)}{2\mJ + 1} 
\mbox{ } . 
\label{Leg-Rec1}
\eeq

\noindent{\bf For useful comparison: ultraspherical harmonics and the Gegenbauer equation} 

\mbox{ } 

\noindent The Gegenbauer alias ultraspherical equation 
\beq
\{1 - X^2\} Y_{,XX} - \{2\lambda + 1\}XY_{,X} + \mJ\{\mJ + 2\lambda\}Y = 0 
\label{Geg}
\eeq
is solved boundedly by the Gegenbauer Polynomials $C_{\sJ}(X;\lambda)$.    
%
%
\noindent The \{k $>$ 3\}-$d$ ultraspherical harmonics equation arising as angular part of higher-$d$ problems is 
straightforwardly separable by the ansatz and change of variables into simple harmonic motion and a sequence of Gegenbauer problems.
\noindent Normalization for these is provided in e.g. \cite{AS, GrRy}.
\noindent The weight function is $\{1 - X^2\}^{\lambda - {1}/{2}}$ between equal-$\lambda$ Gegenbauer polynomials.  
\noindent These furthermore obey the recurrence relation \cite{AS, GrRy} 
\beq
XC_{\sJ}(X;\lambda) = \frac{  \{\mJ + 1\}C_{\sJ + 1}(X; \lambda) + 
\{2\lambda + \mJ - 1\}C_{\sJ - 1}(X; \lambda)}{2\{\mJ + \lambda\}} \mbox{ } . 
\label{Geg-Rec1}
\eeq

\noindent{\bf Tchebychev polynomials of the first kind}

\mbox{ } 

\noindent The Tchebychev polynomials of the first kind T$_{\sn}(x) = \mbox{cos(\mn}\,\mbox{arccos}(x))$ are the solutions of the Tchebychev equation 
\beq
\{1 - x^2\}y_{xx} - x y_x + \mn^2y = 0  \mbox{ } .  
\eeq

\section{Wigner D-functions}\label{Wig-D}

\noindent These are not usually separately tabulated or studied as special functions. 
This is because (see e.g. \cite{Edmonds}) they can be expressed in terms of more basic special functions, which are tabulated and studied as special functions, 
according to the following relations \cite{Edmonds}.   

The Wigner D-function $\mD^{(\sll)}_{\sm\sk}(\alpha, \beta, \gamma)$ solves the equation 
\beq
\{ \pa_{\beta}^2 + \mbox{cot}\,\beta + \mbox{sin}^{-2}\beta\pa_{\alpha}^2 + \pa_{\gamma}^2 - 2\,\mbox{cos}\,\beta\,\pa_{\alpha}\pa_{\gamma}  \} Y 
+ \ml\{\ml + 1\} Y = 0 \mbox{ } . 
\eeq
This separates into two SHM problems and a d-function of $\beta$ alone that maps once again to the hypergeometric equation and thus gives Jacobi polynomials:
\beq
\mD^{(\sll)}_{\sm\sk}(\alpha, \beta, \gamma) = \mbox{exp}(i\nm\gamma)\md^{\sll}_{\sm \sk}(\beta)\mbox{exp}(i\nk\alpha) \mbox{ } , 
\eeq
\beq
\md^{\sll}_{\sm \sk}(\beta) := \sqrt{    \frac{  \{\ml + \nm\}! \{\ml - \nm\}!  }{  \{\ml + \nk\}! \{\ml - \nk\}!  }     }
\mbox{sin}^{\sm - \sk}\mbox{$\frac{\beta}{2}$} \mbox{cos}^{\sm + \sk}\mbox{$\frac{\beta}{2}$}
P^{(\sm - \sk, \sm + \sk)}_{\sk - \sm}(\mbox{cos}\,\beta)  \mbox{ } . 
\eeq
\mbox{ } \mbox{ }  The orthonormality relation for the Wigner D-functions is then
\beq
\frac{1}{8\pi^2}\int_{\alpha = 0}^{2\pi}\int_{\beta = 0}^{\pi}\int_{\gamma = 0}^{2\pi}
\mD^{(\sll_1)*}_{\sm_1\sk_1}(\alpha, \beta, \gamma)
\mD^{(\sll_2)}_{\sm_2\sk_2}         (\alpha, \beta, \gamma)
\d\alpha \, \mbox{sin}\,\beta \, \d\beta \, \d\gamma = 
\frac{1}{2}\sqrt{\frac{\{\ml_1\!+\!\nk_1\}!\{\ml_1\!-\!\nk_1\}!\{\ml_2\!+\!\nk_1\}!\{\ml_2\!-\!\nk_1\}!}
                      {\{\ml_1\!+\!\nm_1\}!\{\ml_1\!-\!\nm_1\}!\{\ml_2\!+\!\nm_1\}!\{\ml_2\!-\!\nm_1\}!}} \delta_{\sm_1\sm_2}\delta_{\sk_1\sk_2} \mbox{ } . 
\eeq

\section{Bessel functions and associated Laguerre polynomials}\label{Bes-Lag}

\noindent These are two families of confluent hypergeometric functions.

\mbox{ } 

\noindent The Bessel equation of order p,
\beq
v^2w_{,vv} + vw_{,v} +\{v^2 - \mp^2\}w = 0 \mbox{ } ,  
\label{Bess}
\eeq
is solved by the Bessel functions.
We denote Bessel functions of the first kind by $\mJ_{\sp}(v)$.  
The family of equations 
\beq
x^2y_{,xx} + \{ 1 - 2\Bigalpha \} xy_{,x} + \big\{ \Bigalpha^2 + \Bigbeta^2\big\{ k^2 x^{2\bigbeta} - \mp^2\big\}\big\}y = 0
\eeq
map to the Bessel equation under the transformations $w = x^{-\bigalpha}y$ and $v = kx^{\bigbeta}$.  
The subcase of this with $\Bigalpha = 1/2$, $\Bigbeta = 1$ and $\mp = \ml + 1/2$ for $\ml \in \mathbb{N}$ are the well-known spherical Bessel functions \cite{AS}.  

\mbox{ } 

\noindent The associated Laguerre polynomials are terminating functions for the confluent hypergeometric o.d.e. 
What we need about them probably is as follows.
\noindent The associated Laguerre equation 
\beq
xy_{,xx} + \{\alpha + 1 - x\}y_{,x} + \mn \, y = 0 \mbox{ }  
\label{Laguerre}
\eeq  
is solved by the associated Laguerre polynomials $\mL_{\sn}^{\alpha}(x)$.  
They obey \cite{AS} the orthogonality relation
\beq
\int_0^{\infty}x^{\alpha}\mbox{exp}(- x)\mL^{\alpha}_{\beta}(x)L^{\alpha}_{\beta^{\prime}}(x)
\d x = 0 \mbox{ unless } \beta = \beta^{\prime} 
\eeq
and the recurrence relation
\beq
x\mL^{\alpha}_{\beta}(x) = \{2\beta + \alpha + 1\}\mL_{\beta}^{\alpha}(x) - 
                               \{\beta + 1\}          \mL_{\beta + 1}^{\alpha}(x) - 
                               \{\beta + \alpha\}     \mL^{\alpha}_{\beta - 1}(x) \mbox{ } .  
\label{LagRec}
\eeq
The 2-$d$ quantum isotropic harmonic oscillator's radial equation for a particle of mass $\mu$ and oscillator frequency $\omega$, 
\beq
-\{{\hbar^2}/{2\mu}\}\{R_{,rr} + {R_{,r}}/{r} + {\nm^2R}/{r^2}\} + {\mu\omega^2r^2R}/{2} = E R \mbox{ } , 
\label{isoho}
\eeq
maps to the associated Laguerre equation under the asymptotically-motivated transformations    
\beq
R = 
\left\{
{\hbar x}/{\mu\omega}
\right\}^{{|\sm|}/{2}}
\mbox{exp}(- x/2)y(x) \mbox{ } , \mbox{ }  
x = {\mu\omega r^2}/{\hbar} \mbox{ } .
\label{isohotrans}
\eeq
This is solved by 
\beq
R \propto r^{|\sm|}\mbox{exp}(\mu\omega r^2/2\hbar)\mL_{\sr}^{|\sm|}
\left(
\mu\omega r^2/{\hbar}
\right)
\label{isohosoln}
\eeq
corresponding to the discrete energies $E = \{|\nm| + 2\nr + 1\}\hbar\omega$ for radial quantum number $\nr \in \mathbb{N}_0$ \cite{Schwinger, Robinett}.


\end{document}